%% file: ss3l2015pub.tex
\newcommand*{\ATLASLATEXPATH}{latex/}
\documentclass[UKenglish,texlive=2011,PAPER]{\ATLASLATEXPATH atlasdoc}
\usepackage[subcaption]{\ATLASLATEXPATH atlaspackage}
\usepackage{\ATLASLATEXPATH atlasprocess}
\usepackage{\ATLASLATEXPATH atlasmisc}
\usepackage{\ATLASLATEXPATH atlasbiblatex}
\usepackage{\ATLASLATEXPATH atlascontribute}
\usepackage{\ATLASLATEXPATH atlasphysics}
\usepackage{\ATLASLATEXPATH atlasbsm}

\graphicspath{{logos/}{FIGURES/}}

\usepackage{ss3l2015pub-defs}
\usepackage[normalem]{ulem}
\usepackage{xparse}
\input{ss3l2015pub-metadata}

\begin{document}

\maketitle

\section{Introduction}
\label{sec:intro}

Supersymmetry (SUSY)~\cite{Golfand:1971iw,Volkov:1973ix,Wess:1974tw,Wess:1974jb,Ferrara:1974pu,Salam:1974ig} is one of the most studied frameworks to extend the Standard Model (SM) beyond the electroweak scale; a general review can be found in Ref.~\cite{Martin:1997ns}. 
In its minimal realisation (MSSM)~\cite{Fayet:1976et,Fayet:1977yc} 
it predicts a new bosonic (fermionic) partner for each fundamental SM fermion (boson), 
as well as an additional Higgs doublet. 
If $R$-parity is conserved~\cite{Farrar:1978xj} the lightest supersymmetric particle (LSP) is stable and is typically the lightest neutralino\footnote{The SUSY partners of the Higgs and electroweak gauge bosons mix to form
the mass eigenstates known as charginos ($\tilde{\chi}^{\pm}_{l}$, $l = 1, 2$ ordered by
increasing mass) and neutralinos ($\tilde{\chi}^{0}_{m}$, $m = 1, \ldots, 4$ ordered by increasing
mass).} \ninoone. 
Only such scenarios are considered in this paper. 
In many models, the LSP can be a viable dark matter candidate~\cite{Goldberg:1983nd,Ellis:1983ew} 
and produce collider signatures with large missing transverse momentum.

In order to address the SM hierarchy problem with SUSY models~\cite{Sakai:1981gr,Dimopoulos:1981yj,Ibanez:1981yh,Dimopoulos:1981zb}, 
TeV-scale masses are required~\cite{Barbieri:1987fn,deCarlos:1993yy} 
for the partners of the gluons (gluinos~\gluino) 
and of the top quark chiral degrees of freedom (top squarks \stopL and \stopR), 
due to the large top Yukawa coupling. 
The latter also favours significant $\stopL$--$\stopR$ mixing, 
so that the lighter mass eigenstate $\stopone$ is in many scenarios lighter than the other squarks~\cite{Inoue:1982pi,Ellis:1983ed}. 
Bottom squarks may also be light, being bound to top squarks by $SU(2)_{\rm L}$ invariance. 
This leads to potentially large production cross-sections 
for gluino pairs ($\gluino\gluino$), top--antitop squark pairs ($\stopone\stoponebar$) and bottom--antibottom squark pairs ($\sbottomone\sbottomonebar$) at the Large Hadron Collider (LHC)~\cite{Borschensky:2014cia}. 
Production of isolated leptons may arise in the cascade decays of those superpartners to SM quarks and neutralinos~\ninoone, 
via intermediate neutralinos~$\tilde\chi^0_{2,3,4}$ or charginos~$\tilde\chi^\pm_{1,2}$ 
that in turn lead to $W$, $Z$ or Higgs bosons, or to lepton superpartners (sleptons). 
Lighter third-generation squarks would also enhance $\gluino\to t\stoponebar$ or $\gluino\to b\sbottomonebar$ branching ratios over the generic decays involving light-flavour squarks,
favouring the production of heavy flavour quarks and, in the case of top quarks, additional leptons. 

This paper 
presents a search for SUSY in final states 
with two leptons (electrons or muons) of the same electric charge (referred to as same-sign (SS) leptons)~\cite{Barnett:1993ea} or three leptons (3L) in any charge combination, 
jets and missing transverse momentum (${\vec p}^{\rm miss}_{\rm T}$, whose magnitude is referred to as \met). 
It is an extension to $\sqrt s=13$ TeV of an earlier search performed by ATLAS with $\sqrt s=8$ TeV data~\cite{paperSS3L}, 
and uses the data collected by the ATLAS experiment~\cite{PERF-2007-01} in proton--proton ($pp$) collisions during 2015. 
Despite the much lower integrated luminosity collected at $\sqrt s=13$~TeV compared to that collected at $\sqrt s=8$~TeV, 
a similar or improved sensitivity at $\sqrt{s}=13$~TeV is expected 
due to the much larger cross-sections predicted for the production of SUSY particles with masses at the TeV scale. 
A similar search for SUSY in this topology was also performed by the CMS Collaboration~\cite{Chatrchyan:2013fea} at $\sqrt s=8$~TeV.
While the same-sign leptons signature is present in many scenarios of physics beyond the SM (BSM), 
SM processes leading to such final states have very small cross-sections. 
Compared to many other BSM searches, analyses based on same-sign leptons therefore allow 
the use of looser kinematic requirements (for example, on \met or the momentum of jets and leptons), 
preserving sensitivity to scenarios with small mass differences between gluinos/squarks and the LSP, or in which $R$-parity is not conserved~\cite{paperSS3L}. 

\begin{figure}[t!]
\centering
\begin{subfigure}[t]{0.24\textwidth}\includegraphics[width=\textwidth]{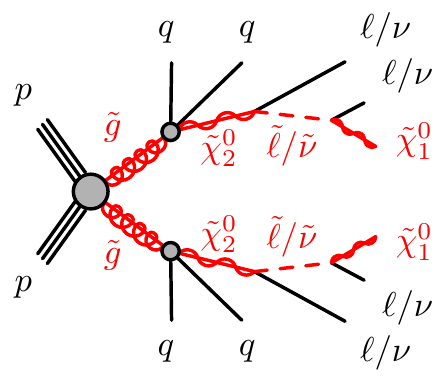}\caption{}\label{fig:feynman_gg2sl}
\end{subfigure}
\begin{subfigure}[t]{0.24\textwidth}\includegraphics[width=\textwidth]{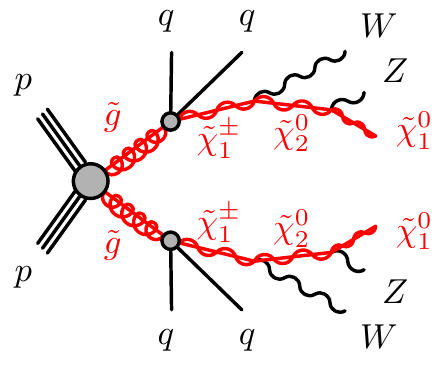}\caption{}\label{fig:feynman_gg2WZ}\end{subfigure}
\begin{subfigure}[t]{0.24\textwidth}\includegraphics[width=\textwidth]{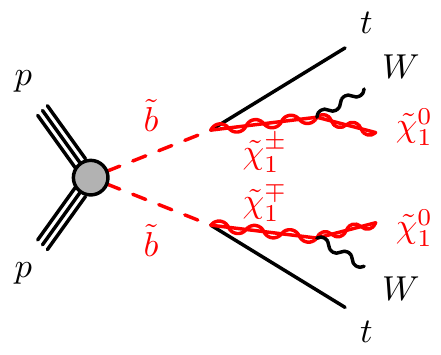}\caption{}\label{fig:feynman_b1b1}\end{subfigure}
\begin{subfigure}[t]{0.24\textwidth}\includegraphics[width=\textwidth]{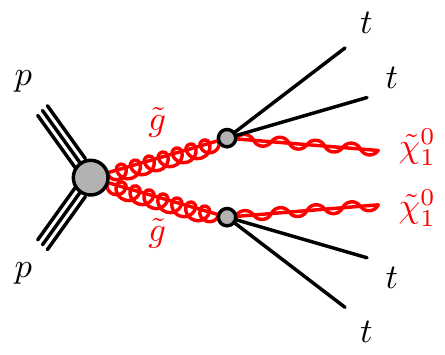}\caption{}\label{fig:feynman_gtt}\end{subfigure}
\caption{SUSY processes featuring gluino (a, b, d) or bottom squark (c) pair production considered in this analysis.}
\label{fig:feynman}
\end{figure}

The sensitivity to a wide range of models is illustrated by the interpretation of the results  
in the context of four different SUSY benchmark processes that may lead to same-sign or three-lepton signatures. 
The first two scenarios focus on gluino pair production with generic decays into light quarks and multiple leptons, 
either involving light sleptons, $\gluino\to q\bar q\ninotwo\to q\bar q \ell\slepton^*\to q\bar q\ell^+\ell^-\ninoone$ (Fig.~\ref{fig:feynman_gg2sl}), 
or mediated by a cascade involving $\chinoonepm$ and $\ninotwo$, 
$\gluino \to q\bar q'\chinoonepm\to q\bar q'W^\pm\ninotwo\to q\bar q'W^\pm Z\ninoone$  (Fig.~\ref{fig:feynman_gg2WZ}). 
The other two scenarios are motivated by the expectation that the third-generation squarks are lighter than the other squarks 
and target the direct production of $\sbottomone\sbottomonebar$ pairs 
with subsequent chargino-mediated $\sbottomone\to tW^-\ninoone$ decays (Fig.~\ref{fig:feynman_b1b1})  
or the production of $\gluino\gluino$ pairs decaying as $\gluino\to t\bar t\ninoone$ via an off-shell top squark (Fig.~\ref{fig:feynman_gtt}). 

Four signal regions (SRs) are designed to achieve good sensitivity for these SUSY scenarios, mainly characterised by the number of $b$-tagged jets or reconstructed leptons. 
They are detailed in Section~\ref{sec:selection}, preceded by descriptions of the experimental apparatus~(Section~\ref{sec:detector}) and
the simulated samples~(Section~\ref{sec:dataMC}).
Section~\ref{sec:bkg} is devoted to the estimation of the
contribution from SM processes to the signal regions, 
validated by comparisons with data in dedicated regions. 
The results are presented in Section~\ref{sec:result}, 
together with the statistical tests used to interpret the results in the context of the SUSY benchmark scenarios. 
Finally, Section~\ref{sec:conclusion} summarises the main conclusions of this paper.

\section{The ATLAS detector}
\label{sec:detector}

The ATLAS experiment~\cite{PERF-2007-01} is a multi-purpose particle detector with a forward-backward symmetric cylindrical
geometry and nearly $4\pi$ coverage in solid angle.\footnote{ATLAS uses
  a right-handed coordinate system with its origin at the nominal
  interaction point (IP) in the centre of the detector and the
  $z$-axis along the beam pipe. The $x$-axis points from the IP to the
  centre of the LHC ring, and the $y$-axis points upward. Cylindrical
  coordinates ($r$, $\phi$) are used in the transverse plane, $\phi$
  being the azimuthal angle around the beam pipe. The pseudorapidity
  is defined in terms of the polar angle $\theta$ as $\eta = -\ln
  \tan(\theta/2)$. Rapidity is defined as $y=0.5 \ln\left[(E + p_z )/(E - p_z )\right]$ where $E$ denotes the energy and $p_z$ is the component of the momentum along the beam direction.}
The interaction point is surrounded by an inner detector (ID), a
calorimeter system, and a muon spectrometer.

The ID provides precision tracking of charged particles for
pseudorapidities $|\eta| < 2.5$ and is surrounded by a superconducting solenoid providing a \SI{2}{T} axial magnetic field.
It consists of pixel and silicon-microstrip detectors inside a
transition radiation tracker.  
One significant upgrade for the $\sqrt{s}=13$~TeV running period is the presence of the
Insertable B-Layer~\cite{CERN-LHCC-2010-013}, an additional pixel layer close to the interaction point, which 
provides high-resolution hits at small radius to improve the tracking performance.

In the pseudorapidity region $|\eta| < 3.2$, high-granularity lead/liquid-argon (LAr)
electromagnetic (EM) sampling calorimeters are used.
A steel/scintillator tile calorimeter measures hadron energies for
$|\eta| < 1.7$.
The endcap and forward regions, spanning $1.5<|\eta| <4.9$, are 
instrumented with LAr calorimeters for both the EM and hadronic
measurements.

The muon spectrometer consists of three large superconducting toroids
with eight coils each, 
a system of trigger and precision-tracking chambers, 
which provide triggering and tracking capabilities in the
ranges $|\eta| < 2.4$ and $|\eta| < 2.7$, respectively.

A two-level trigger system is used to select events. The first-level
trigger is implemented in hardware and uses a subset of the detector
information. This is followed by the software-based High-Level Trigger stage,
which can run offline reconstruction and
calibration software, reducing the event rate to about \SI{1}{kHz}.

\section{Dataset and simulated event samples}
\label{sec:dataMC}

\input{dataMC}

\section{Event selection}
\label{sec:selection}

\input{sel.tex}

\section{Background estimation}
\label{sec:bkg}

\input{bkg}

\section{Results}
\label{sec:result}

\input{results}

\FloatBarrier

\section{Conclusion}
\label{sec:conclusion}

A search for supersymmetry in events with exactly two same-sign leptons or at least three leptons, multiple jets, 
$b$-jets and $\met$ is presented. 
The analysis is performed with proton--proton collision data at $\sqrt{s}=13\TeV$ collected with the ATLAS detector at the Large Hadron Collider corresponding to an integrated luminosity of 3.2~fb$^{-1}$. 
With no significant excess over the Standard Model expectation observed,
results are interpreted in the framework of simplified models featuring gluino and bottom squark production.
In the $\gluino\gluino$ simplified models considered, $m_{\gluino}\lesssim 1.1$--$1.3~\TeV$ and $m_{\ninoone}\lesssim 550$--$850~\GeV$ 
are excluded at 95\% confidence level depending on the model parameters. Bottom squark masses of $m_{\sbottomone}\lesssim 540~\GeV$ 
are also excluded for a light $\ninoone$ in a $\sbottomone\sbottomonebar$ simplified model with $\sbottomone\to tW^-\ninoone$.
These results are complementary to those of previous searches and extend the exclusion limits they set.

\section*{Acknowledgements}

\input{Acknowledgements}

\printbibliography

\clearpage

\newpage \input{atlas_authlist}

\end{document}

%% file: ss3l2015pub-metadata.tex
\AtlasTitle{Search for supersymmetry at $\sqrt{s}=13$~TeV in final states with jets and two same-sign leptons or three leptons with the ATLAS detector}

\author{The ATLAS Collaboration}

\PreprintIdNumber{CERN-EP-2016-027}

\arXivId{1602.09058}

\HepDataRecord{http://hepdata.cedar.ac.uk/view/ins1424844}

\AtlasJournalRef{Eur.\ Phys.\ J.\ C (2016) 76:259}
\AtlasDOI{10.1140/epjc/s10052-016-4095-8}

\AtlasAbstract{A search for strongly produced supersymmetric particles is conducted using signatures involving multiple energetic jets and either two isolated leptons ($e$ or $\mu$) with the same electric charge or at least three isolated leptons. The search also utilises $b$-tagged jets, missing transverse momentum and other observables to extend its sensitivity. The analysis uses a data sample of proton--proton collisions at $\sqrt{s}= 13$~TeV recorded with the ATLAS detector at the Large Hadron Collider in 2015 corresponding to a total integrated luminosity of 3.2~fb$^{-1}$. No significant excess over the Standard Model expectation is observed.  
The results are interpreted in several simplified supersymmetric models and extend the exclusion limits from previous searches.
In the context of exclusive production and simplified decay modes, gluino masses are excluded at $95\%$ confidence level 
up to 1.1--1.3~TeV for light neutralinos (depending on the decay channel), 
and bottom squark masses are also excluded up to 540 GeV. 
In the former scenarios, neutralino masses are also excluded up to 550-850 GeV for gluino masses around 1 TeV. 
}

%% file: dataMC.tex
The data were collected by the ATLAS detector during 2015 with a peak 
instantaneous luminosity of $L=5.2\times 10^{33}$~cm$^{-2}$s$^{-1}$, 
a bunch spacing of 25~ns, and a mean number of additional $pp$ interactions per bunch crossing 
(pile-up) in the dataset of $\langle \mu \rangle = 14$.
After the application of beam, detector and data quality requirements, 
the integrated luminosity considered in this analysis corresponds to 3.2~fb$^{-1}$ with an uncertainty of $\pm5\%$. 
The luminosity and its uncertainty are derived following a methodology similar to that detailed in
Ref.~\cite{DAPR-2011-01} from a preliminary calibration of the
luminosity scale using a pair of $x$--$y$ beam separation scans performed in
August 2015.

Monte Carlo (MC) simulated event samples are used to aid in the
estimation of the background from SM processes and to model the SUSY signal. 
The MC samples are processed through an ATLAS detector simulation~\cite{Aad:2010ah} based on 
{\sc Geant4}~\cite{Agostinelli:2002hh} 
or a fast simulation using a parameterisation of the calorimeter response 
and {\sc Geant4} for the other parts of the detector~\cite{ATL-PHYS-PUB-2010-013}   
and are reconstructed in the same manner as the data. 

Diboson processes with four charged leptons ($\ell$), three charged leptons and one neutrino, or 
two charged leptons and two neutrinos are simulated using the \SHERPA~v2.1.1 generator~\cite{Gleisberg:2008ta}, 
and are described in detail in Ref.~\cite{pubnote_mc_multiboson}. 
The matrix elements contain the doubly resonant $WW$, $WZ$ and $ZZ$ processes and all other diagrams with
four or six electroweak vertices (such as same-electric-charge $W$ boson production in association with two jets, $W^\pm W^\pm jj$). 
Fully leptonic triboson processes ($WWW$, $WWZ$, $WZZ$ and $ZZZ$) with up to six charged leptons are also simulated using \SHERPA~v2.1.1 
and described in Ref.~\cite{pubnote_mc_multiboson}. 
The $4\ell$ and $2\ell+2\nu$ processes are calculated at next-to-leading order (NLO) for up to one additional parton; 
final states with two and three additional partons are calculated at leading order (LO). 
The $WWZ\to 4\ell+2\nu$ or $2\ell+4\nu$ processes are calculated at LO with up to two additional partons. 
The $3\ell +1\nu$ process is calculated at 
NLO and up to three extra partons at LO using the Comix~\cite{Gleisberg:2008fv} 
and OpenLoops~\cite{Cascioli:2011va} 
matrix element generators and merged with the \SHERPA parton shower~\cite{Schumann:2007mg} 
using the ME+PS@NLO prescription~\cite{Hoeche:2012yf}. 
The $WWW/WZZ\to 3\ell+3\nu$, $WZZ\to 5\ell+1\nu$, $ZZZ\to 6\ell+0\nu$, $4\ell+2\nu$ or $2\ell+4\nu$ processes 
are calculated with the same configuration but with up to only two extra partons at LO. 
The CT10~\cite{Lai:2010vv} parton distribution function (PDF) set is used for all \SHERPA samples in conjunction with 
a dedicated tuning of the parton shower parameters developed by the \SHERPA authors. 
The generator cross-sections (at NLO for most of the processes) are used when normalising these backgrounds.

Samples of $t\bar{t} V$ (with $V=W$ and $Z$, including non-resonant $Z/\gamma^*$ contributions) 
and $t\bar{t}WW$ production are generated at LO with \MADGRAPH v2.2.2~\cite{Alwall:2011uj} interfaced to the \PYTHIA 8.186~\cite{Sjostrand:2007gs} parton shower model,
with up to two ($\ttbar W$), one ($\ttbar Z$) or no ($\ttbar WW$) extra partons included in the matrix element; 
they are described in detail in Ref.~\cite{pubnote_mc_ttv}. 
\MADGRAPH is also used to simulate the $tZ$, $\ttbar\ttbar$ and $\ttbar t$ processes. 
The {\sc A14} set of tuned parameters (tune)~\cite{pub-2014-021} is used together with the {\sc NNPDF23LO} PDF set~\cite{Ball:2012cx}. 
The $\ttbar W$, $\ttbar Z$, $\ttbar WW$ and $\ttbar\ttbar$ events are normalised to their NLO cross-section~\cite{Alwall:2014hca} 
while the generator cross-section is used for $tZ$ and $\ttbar t$.

Production of a Higgs boson in association with a $\ttbar$ pair is simulated using \AMCATNLO~\cite{Alwall:2014hca} 
(in \MADGRAPH v2.2.2) interfaced to \HERWIG 2.7.1~\cite{Corcella:2000bw}.  
The UEEE5 underlying-event tune is used together with the CTEQ6L1~\cite{Pumplin:2002vw} (matrix element) and CT10~\cite{Lai:2010vv} (parton shower) PDF sets.
Simulated samples of SM Higgs boson production in association with a $W$ or $Z$ boson are produced with \PYTHIA 8.186, using the {\sc A14} tune and the {\sc NNPDF23LO} PDF set. 
Events are normalised with cross-sections calculated at NLO~\cite{Dittmaier:2012vm}.

The signal SUSY processes are generated from LO matrix elements with up to two extra partons, 
using the \MADGRAPH v2.2.3 generator interfaced to \PYTHIA 8.186 with the {\sc A14} tune for the modelling of the SUSY decay chain, 
parton showering, hadronisation and the description of the underlying event. 
Parton luminosities are provided by the {\sc NNPDF23LO} PDF set. 
Jet--parton matching is realised following the CKKW-L prescription~\cite{Lonnblad:2011xx}, with a matching scale set to one quarter of the pair-produced superpartner mass. 
Signal cross-sections are calculated to NLO in the strong coupling constant, 
adding the resummation of soft gluon emission at next-to-leading-logarithmic accuracy (NLO+NLL)~\cite{Beenakker:1996ch,Kulesza:2008jb,Kulesza:2009kq,Beenakker:2009ha,Beenakker:2011fu}. 
The nominal cross-section and the uncertainty are taken from an envelope of cross-section predictions using different PDF sets and factorisation and renormalisation scales, 
as described in Ref.~\cite{Kramer:2012bx}. 
The production cross-section of gluino pairs with a mass of \SI{1.2}{\TeV} is \SI{86}{\fb} at $\sqrt{s}=13$~TeV (compared with \SI{4.4}{\fb} at $\sqrt{s}=8$~TeV), 
while the production cross-section of pairs of bottom squarks with a mass of \SI{500}{\GeV} is \SI{520}{\fb} at $\sqrt{s}=13$~TeV (compared with \SI{86}{\fb} at $\sqrt{s}=8$~TeV). 

In all MC samples, except those produced by \SHERPA, the {\sc EvtGen}~v1.2.0 program~\cite{EvtGen} is used to model the properties of the bottom and charm hadron decays. 
To simulate the effects of additional $pp$ collisions in the same and nearby bunch crossings, 
additional interactions are generated using the soft QCD processes of \PYTHIA 8.186 
with the A2 tune~\cite{ATLAS-PHYS-PUB-2012-003} and the MSTW2008LO PDF~\cite{Martin:2009iq}, 
and overlaid onto the simulated hard scatter event. 
The Monte Carlo samples are reweighted so that the distribution of the number of reconstructed vertices matches the distribution observed in the data.

%% file: sel.tex
Candidate events are required to have a reconstructed vertex~\cite{ATL-PHYS-PUB-2015-026}, 
with at least two associated tracks with $\pt >400$~MeV, 
and the vertex with the highest sum of squared transverse 
momentum of the tracks is considered as primary vertex.
In order to perform background estimations using data, 
two categories of electrons and muons are defined: 
``candidate'' and ``signal'' (the latter being a subset of the ``candidate'' leptons satisfying tighter selection criteria). 

Electron candidates are reconstructed from an isolated electromagnetic 
calorimeter energy deposit matched to an ID track and are required to have $|\eta|<2.47$, 
a transverse momentum $\pT>\SI{10}{\GeV}$, 
and to pass a loose likelihood-based identification requirement~\cite{elecperf,ATL-PHYS-PUB-2015-041}. 
The likelihood input variables include measurements of calorimeter shower shapes and measurements of track properties from the ID. 
Candidates within the transition region between the barrel and endcap electromagnetic calorimeters,
$1.37<|\eta|<1.52$, are removed. 
The track matched with the electron must have a significance of the transverse impact parameter 
with respect to the reconstructed primary vertex, $d_0$, of $\vert d_0\vert/\sigma(d_0) < 5$.

Muon candidates are reconstructed in the region $|\eta|<2.5$ 
from muon spectrometer tracks matching ID tracks.
All muons must have $\pT>\SI{10}{\GeV}$ and must pass the medium identification requirements defined in Ref.~\cite{Aad:2016jkr}, 
based on selections on the number of hits in the different ID and muon spectrometer subsystems, 
and the significance of the charge to momentum ratio $q/p$~\cite{Aad:2016jkr}.

Jets are reconstructed with the anti-$k_t$
algorithm~\cite{Cacciari:2008} with radius parameter $R=0.4$, using three-dimensional energy
clusters in the calorimeter~\cite{caloclusters} as input. 
All jets must have $\pT>\SI{20}{\GeV}$ and $|\eta|<2.8$.
Jets are calibrated as described in Ref.~\cite{ATLAS-PHYS-PUB-2015-015}.
In order to reduce the effects of pile-up, 
for jets with $\pt<\SI{50}{GeV}$ and $|\eta|<2.4$ a significant fraction of the tracks associated with each jet
must have an origin compatible with the primary vertex, 
as defined by the jet vertex tagger~\cite{ATLAS-CONF-2014-018}. 
Furthermore, for all jets the expected average energy contribution from
pile-up clusters is subtracted according to the jet area~\cite{ATLAS-PHYS-PUB-2015-015}.

Identification of jets containing $b$-hadrons ($b$-tagging) is performed with the MV2c20 algorithm, 
a multivariate discriminant making use of track impact parameters 
and reconstructed secondary vertices~\cite{Aad:2015ydr,ATL-PHYS-PUB-2015-022}.
A requirement is chosen corresponding to a 70\% average efficiency 
obtained for $b$-jets in simulated $\ttbar$ events. 
The rejection factors for light-quark jets, $c$-quark jets and hadronically decaying $\tau$ leptons in simulated $\ttbar$ events 
are approximately 440, 8 and 26, respectively~\cite{ATL-PHYS-PUB-2015-022}. 
Jets with $|\eta|<2.5$ which satisfy this $b$-tagging requirement are identified as $b$-jets. 
To compensate for differences between data and MC simulation in the $b$-tagging efficiencies and mis-tag rates, 
correction factors are applied to the simulated samples~\cite{ATL-PHYS-PUB-2015-022}. 

After object identification, overlaps between objects are resolved. 
Any jet within a distance $\Delta R_y =\sqrt{(\Delta y)^2+(\Delta\phi)^2}$ = 0.2 of an electron candidate is discarded, 
unless the jet has a value of the MV2c20 discriminant larger than the value corresponding to approximately an 80\% $b$-tagging efficiency, 
in which case the electron is discarded since it is likely originating from a semileptonic $b$-hadron decay. 
Any remaining electron within $\Delta R_y=$ 0.4 of a jet is discarded. 
Muons within $\Delta R_y=$ 0.4 of a jet are also removed. 
However, if the jet has fewer than three associated tracks, the muon is kept and the jet is discarded instead 
to avoid inefficiencies for high-energy muons undergoing significant energy loss in the calorimeter.

Signal electrons must satisfy a tight likelihood-based identification requirement~\cite{elecperf,ATL-PHYS-PUB-2015-041} 
and have $|\eta|<2$ to reduce the impact of electron charge mis-identification. 
Signal muons must fulfil the requirement of $\vert d_0\vert/\sigma(d_0) < 3$. 
The track associated to the signal leptons must have a longitudinal impact parameter with respect to the reconstructed primary vertex, $z_0$, 
satisfying $\vert z_0 \sin\theta\vert  < 0.5$~mm. 
Isolation requirements are applied to both the signal electrons and muons. 
The  scalar sum of the \pt of tracks within a variable-size cone around the lepton, 
excluding its own track, must be less than 6\% of the lepton \pt. 
The track isolation cone radius for electrons (muons) 
$\Delta R_\eta=\sqrt{(\Delta\eta)^2+(\Delta\phi)^2}$ 
is given by 
the smaller of $\Delta R_\eta = \SI{10}{~GeV}/\pt$ and $\Delta R_\eta = 0.2\,(0.3)$, 
that is, a cone of size $0.2\,(0.3)$ at low $\pt$ but narrower for high-\pt leptons. 
In addition, in the case of electrons the energy of calorimeter energy clusters in a cone of $\Delta R_\eta = 0.2$ around the electron 
(excluding the deposition from the electron itself) must be less than 6\% of the electron \pt. 
Simulated events are corrected to account for minor differences in the lepton trigger, reconstruction and 
identification efficiencies between data and MC simulation.

The missing transverse momentum ${\vec p}^{\rm miss}_{\rm T}$ is defined as the negative vector sum of the transverse momenta of all identified physics objects (electrons, photons, muons, jets) and an additional soft term. The soft term is constructed from all tracks that are not associated with any physics object, and that are associated with the primary vertex. In this way, the $\met$ is adjusted for the best calibration of the jets and the other identified physics objects above, while maintaining pile-up independence in the soft term~\cite{ATL-PHYS-PUB-2015-027, ATL-PHYS-PUB-2015-023}.

Events are selected using a combination (logical OR) of dilepton and $\met$ triggers, 
the latter being used only for events with $\met>\SI{250}{GeV}$. 
The trigger-level requirements on $\met$ and the leading and subleading lepton \pt are looser than those applied offline 
to ensure that trigger efficiencies are constant in the relevant phase space. 
Events of interest are selected if they contain at least two signal leptons with $\pt>20$~\GeV . 
If the event contains exactly two signal leptons, they are required to have the same electric charge. 

To maximise the sensitivity in different signal models, 
four overlapping signal regions are defined as shown in Table~\ref{tab:SRdef3}, with requirements on the number of signal leptons ($N_{\rm{lept}}^{\rm{signal}}$), 
the number of $b$-jets with $\pt>\SI{20}{\GeV}$ ($N_{b\rm{-jets}}^{20}$), 
the number of jets with $\pt>\SI{50}{\GeV}$ regardless of their flavour ($N_{\rm{jets}}^{50}$), 
\met\ 
and the effective mass (\meff), defined as the scalar sum of the $\pt$ of the signal leptons and jets (regardless of their flavour) in the event plus the \met. 

\begin{table}[tbh!]
\caption{Summary of the event selection criteria for the signal regions (see text for details).}
\hspace{0.5cm}
\def\arraystretch{1.3}
\label{tab:SRdef3}
\centering
\begin{tabular}{c|c|c|c|c|c}
\hline 
\hline
Signal region  &  $N_{\rm{lept}}^{\rm{signal}}$   & $N_{b\rm{-jets}}^{20}$    & $N_{\rm{jets}}^{50}$  & \met\ [GeV] & \meff\ [GeV]   \\
\hline\hline
SR0b3j &  $\ge$3  &    $=$0 &  $\ge$3 &  $>$200 & $>$550 \\
SR0b5j &  $\ge$2  &    $=$0 &  $\ge$5 &  $>$125 & $>$650 \\
SR1b     &  $\ge$2  &    $\ge$1  &  $\ge$4 &  $>$150 & $>$550 \\
SR3b     &   $\ge$2  &   $\ge$3  &  - & $>$125 & $>$650   \\
\hline\hline
\end{tabular}
\end{table}

Each signal region is motivated by a different SUSY scenario. 
The SR0b3j and SR0b5j signal regions are sensitive to gluino-mediated and directly produced squarks of the first and second generations 
leading to final states particularly rich in leptons (Fig.~\ref{fig:feynman_gg2sl}) or in jets (Fig.~\ref{fig:feynman_gg2WZ}), 
but with no enhancement of the production of $b$-quarks. 
Third-generation squark models resulting in final states with two $b$-quarks, 
such as direct bottom squark production (Fig.~\ref{fig:feynman_b1b1}), are targeted by the SR1b signal region. 
Finally, the signal region SR3b targets gluino-mediated top squark production resulting in final states with four $b$-quarks (Fig.~\ref{fig:feynman_gtt}).

The values of acceptance times efficiency of the SR selections for the SUSY signal models 
in Fig.~\ref{fig:feynman} typically range between 1\% and 6\% for $m_{\gluino}=\SI{1.2}{TeV}$ or $m_{\sbottomone}=\SI{600}{GeV}$, and a light \ninoone.

%% file: bkg.tex
Three main sources of SM background can be distinguished in this analysis. 
A first category consists of events with two same-sign prompt leptons or at least three prompt leptons, 
mainly from $\ttbar V$ and diboson processes. 
Other types of background events include those containing electrons with mis-measured charge, mainly from the production of top quark pairs, 
and those containing at least one non-prompt or fake lepton, 
which mainly originate from hadron decays in events containing top quarks or of $W$ bosons in association with jets. 

\subsection{Background estimation methods} 

The estimation of the SM background processes with two same-sign prompt leptons or at least three prompt leptons 
is performed using the MC samples described in Section~\ref{sec:dataMC}. 
Since diboson and $\ttbar V$ events are the main backgrounds in the signal regions, 
dedicated validation regions with an enhanced contribution from these processes 
are defined to verify the background predictions (see Section~\ref{sec:valid}).

Background events due to charge mis-identification, dominated by electrons having emitted 
a hard brems\-strah\-lung photon which subsequently converted to an electron--positron pair, 
are referred to as “charge-flip”. 
The probability of mis-identifying the charge of a muon is checked in both data and MC simulation, 
and found to be negligible in the kinematic range relevant to this analysis.
The contribution of charge-flip events is estimated using data. 
The electron charge-flip probability is extracted in a $Z/\gamma^{*}\to ee$ data sample using a likelihood fit 
which takes as input the numbers of same-sign and opposite-sign electron pairs observed in the sample. 
The charge-flip probability is a free parameter of the fit and is extracted as a function of the electron $\pt$ and $\eta$. 
The event yield of this background in the signal or validation regions is obtained by applying the measured charge-flip probability 
to data regions with the same kinematic requirements as the signal or validation regions but with opposite-sign lepton pairs.

The contribution from fake or non-prompt (FNP) leptons (such as hadrons mis-identified as leptons, 
leptons originating from heavy-flavour decays, and electrons from photon conversions) 
is also estimated from data with a matrix method similar to that described in Ref.~\cite{paperSS3L}. 
In this method, two types of lepton identification criteria are defined: ``tight'', 
corresponding to the signal lepton criteria described in Section~\ref{sec:selection}, 
and ``loose'', corresponding to candidate leptons. 
The matrix method relates the number of events containing prompt or FNP leptons 
to the number of observed events with tight or loose-not-tight leptons 
using the probability for loose prompt or FNP leptons to satisfy the tight criteria. 
The probability for loose prompt leptons to satisfy the tight selection criteria 
is obtained using a $Z/\gamma^*\to\ell\ell$ data sample 
and is modelled as a function of the lepton $\pt$ and $\eta$. 
The probability for loose FNP leptons to satisfy the tight selection criteria is determined from data 
in a SS control region enriched in non-prompt leptons originating from heavy-flavour decays. 
This region contains events with at least one $b$-jet, 
one tight muon with $\pt > \SI{40}{GeV}$ (likely prompt) and an additional loose electron or muon (likely FNP).
The contribution from prompt leptons and charge mis-measured electrons to this region is subtracted from the observed event yields.

The data-driven background estimates are cross-checked with an MC-based technique. 
In this method, the contributions from processes with FNP leptons and electron charge mis-identification 
are obtained from MC simulation and normalised to data in dedicated control regions at low jet multiplicity, low $\met$, and 
either with or without $b$-jets. 
The normalisation is performed using five multipliers: one to correct 
the electron charge mis-identification rate, and four to correct the contributions from FNP 
electrons or muons originating from $b$-jets or light-flavour jets, respectively.
In addition to the MC samples listed in Section~\ref{sec:dataMC}, this method employs samples of top quark pair production generated with the \POWHEG-Box v2 generator interfaced to \PYTHIA 6.428~\cite{Sjostrand:2006za}, 
as well as samples of simulated $W$+jets and $Z$+jets events generated with \POWHEG-Box v2 interfaced to \PYTHIA 8.186.

\subsection{Systematic uncertainties on the background estimation}

Table~\ref{tab:SR_syst} summarises the contributions of the different sources of systematic uncertainty 
in the total SM background predictions in the signal regions.

The systematic uncertainties related to the same-sign prompt leptons background estimation 
arise from the accuracy of the theoretical and experimental modelling in the MC simulation.
The primary sources of systematic uncertainties are related to
the jet energy scale calibration, 
the jet energy resolution, 
$b$-tagging efficiency, 
and MC modelling and theoretical cross-section uncertainties. 
The cross-sections used to normalise the MC samples are varied according to the uncertainty in the 
cross-section calculation, that is, 6\% for diboson, 13\% for $\ttbar W$ and 12\% $\ttbar Z$ production~\cite{Alwall:2014hca}.
Additional uncertainties are assigned to these backgrounds 
to account for the modelling of the kinematic distributions in the MC simulation. 
For $\ttbar W$ and $\ttbar Z$, the predictions from the \MADGRAPH and \SHERPA generators are compared, 
leading to a $\sim$30\% uncertainty for these processes after the SR selections. 
For dibosons, uncertainties are estimated by varying the renormalisation, factorisation and resummation scales used to generate these samples, 
leading to a $\sim$30\% uncertainty for these processes after the SR selections. 
For triboson, $\ttbar h$, $\ttbar \ttbar$ and $tZ$ production processes, which constitute a small background in all signal regions, a 50\% uncertainty on the event yields is assumed.

Uncertainties in the FNP lepton background estimate are assigned due to the limited number of data events with loose and tight leptons.
In addition, systematic uncertainties of 50--60\% are assigned to the probabilities for loose FNP leptons to satisfy the tight signal criteria to account for potentially different FNP compositions (heavy flavour, light flavour or conversions) between the regions used to measure these probabilities and the SRs, as well as the contamination from prompt leptons in the former regions.
This leads to overall FNP background uncertainties in the total background estimates of 18--21\% depending on the signal region.

For the charge-flip background prediction, the main uncertainties originate from the statistical 
uncertainty of the charge-flip probability measurements and the background contamination of the 
sample used to extract the charge-flip probability. 

\begin{table}[h!]
\begin{center}
\caption{The main sources of systematic uncertainty on the SM background estimates for the four signal regions are shown 
and their values given as relative uncertainties in the expected signal region background event yields. 
The individual components can be correlated and therefore do not necessarily add up in quadrature to the total systematic uncertainty.
For reference, the total number of expected background events is also shown.
}
\label{tab:SR_syst}
{\small
\begin{tabular}{lrrrr}
\noalign{\smallskip}\hline\hline\noalign{\smallskip}
         & SR0b3j         & SR0b5j     & SR1b & SR3b     \\[-0.05cm]
\noalign{\smallskip}\hline\hline\noalign{\smallskip}
Diboson theoretical uncertainties    & 23\%  &  16\%   &  1\%  &$<$1\%   \\
$\ttbar V$ theoretical uncertainties & 3\%   &  4\%    & 13\%  &  9\%   \\
Other theoretical uncertainties      & 5\%   &  3\%    &  9\%  & 15\%   \\
\noalign{\smallskip}\hline\noalign{\smallskip}
MC statistical uncertainties         & 11\%  &  14\%   &  3\%  &  6\%   \\
\noalign{\smallskip}\hline\noalign{\smallskip}
Jet energy scale        & 12\%   &  11\%  & 6\%    & 5\%   \\
Jet energy resolution   & 3\%    &  9\%   & 2\%    & 3\%   \\
$b$-tagging             & 4\%    &  6\%   & 3\%    & 10\%   \\
PDF                     & 6\%    &  6\%   & 6\%    & 8\%   \\
Fake/non-prompt leptons & 18\%    &  20\%   & 18\%   & 21\%   \\
Charge flip             & --     & 1\% & 3\%    & 8\%   \\
\noalign{\smallskip}\hline\noalign{\smallskip}
Total background uncertainties & 30\%   & 34\%   & 22\%   & 31\%   \\
\noalign{\smallskip}\hline\hline\noalign{\smallskip}
Total background events & $1.5$ & $0.88$ & $4.5$ & $0.80$\\
\noalign{\smallskip}\hline\hline\noalign{\smallskip}
\end{tabular}
}
\end{center}
\end{table}

\subsection{Validation of background estimates}
\label{sec:valid}

\begin{figure}[p!]
\centering
\begin{subfigure}[t]{0.46\textwidth}\includegraphics[width=\textwidth]{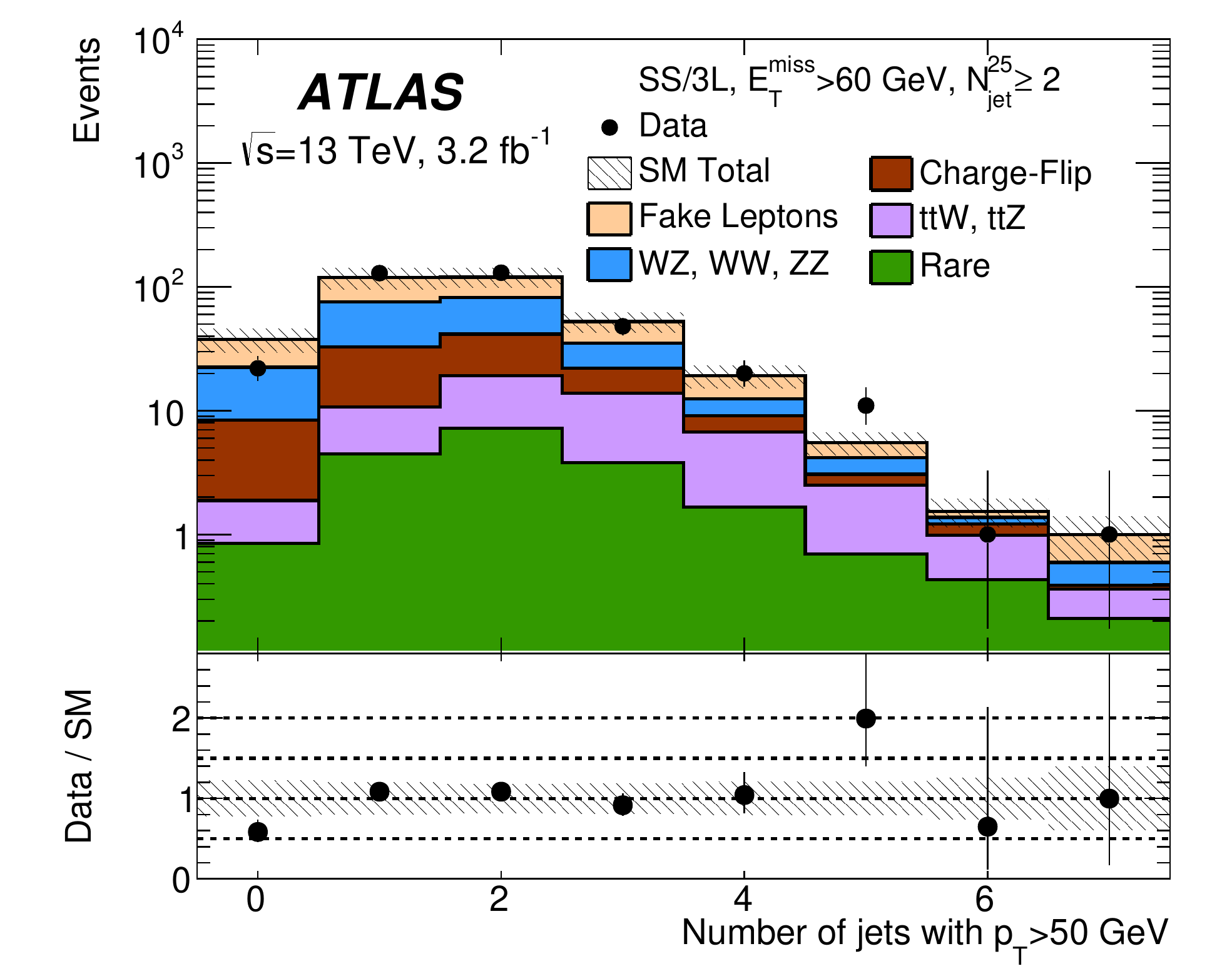}
\caption{}\label{fig:VRa}\end{subfigure}
\begin{subfigure}[t]{0.46\textwidth}\includegraphics[width=\textwidth]{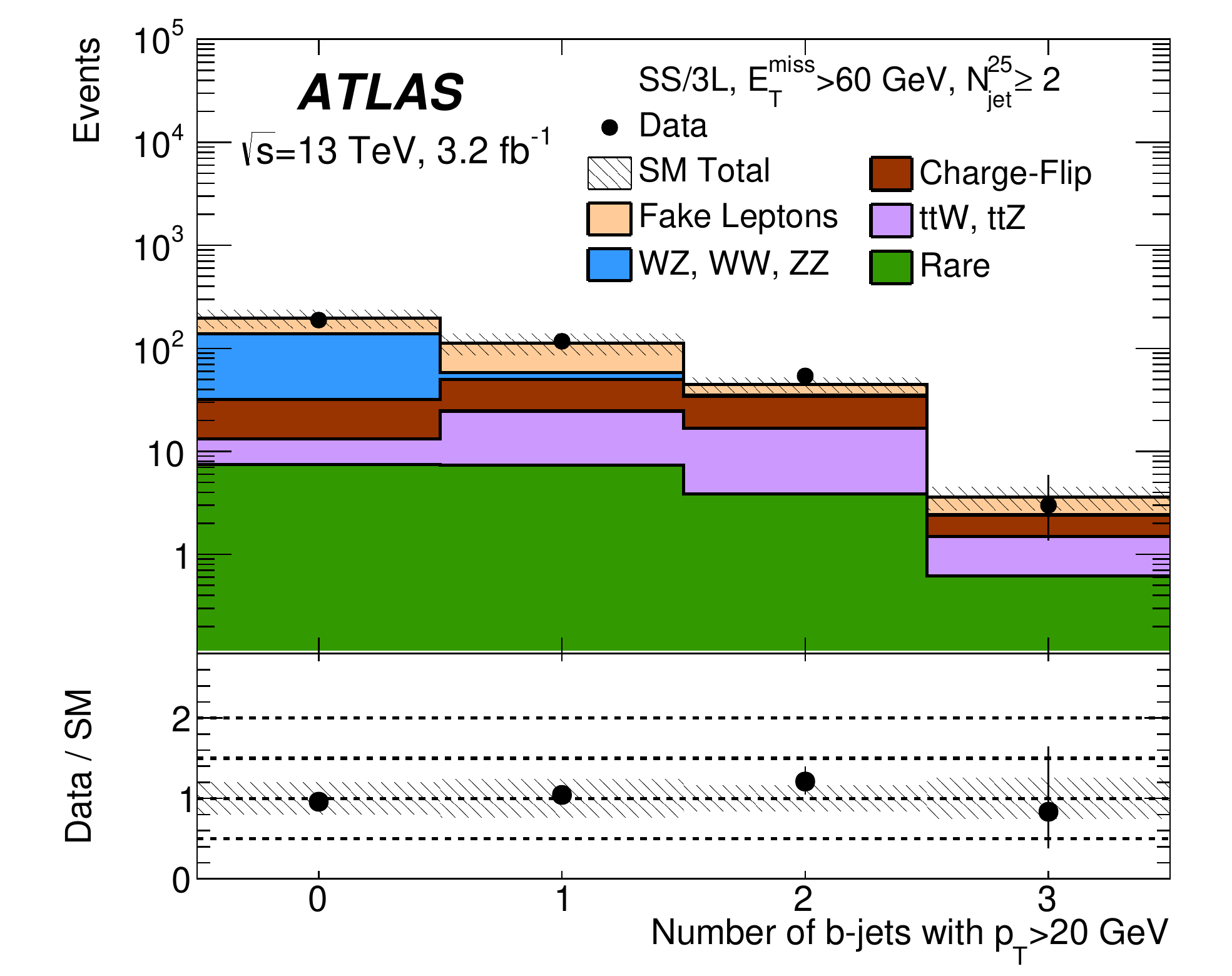}
\caption{}\label{fig:VRb}\end{subfigure}
\begin{subfigure}[t]{0.46\textwidth}\includegraphics[width=\textwidth]{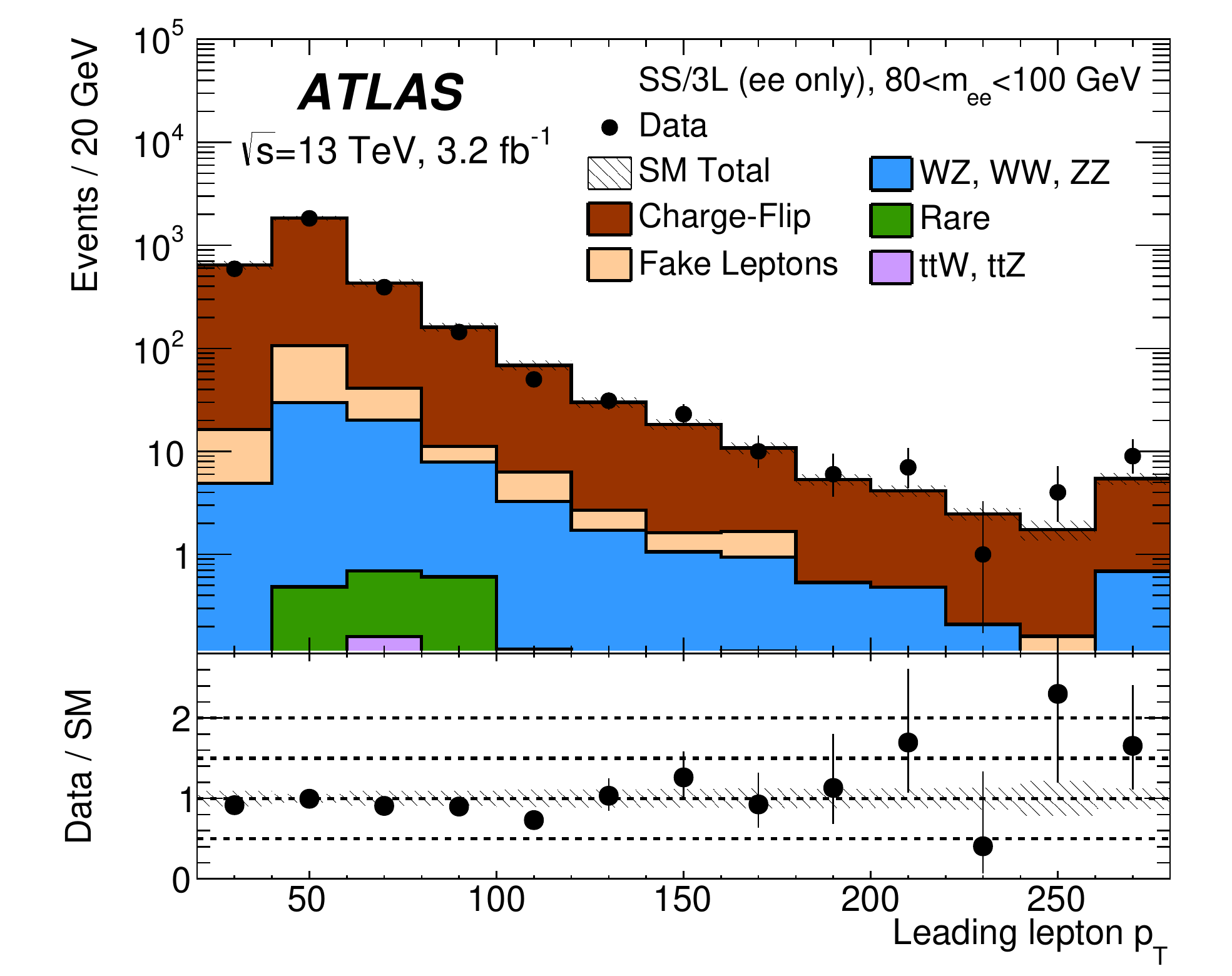}
\caption{}\label{fig:VRc}\end{subfigure}
\begin{subfigure}[t]{0.46\textwidth}\includegraphics[width=\textwidth]{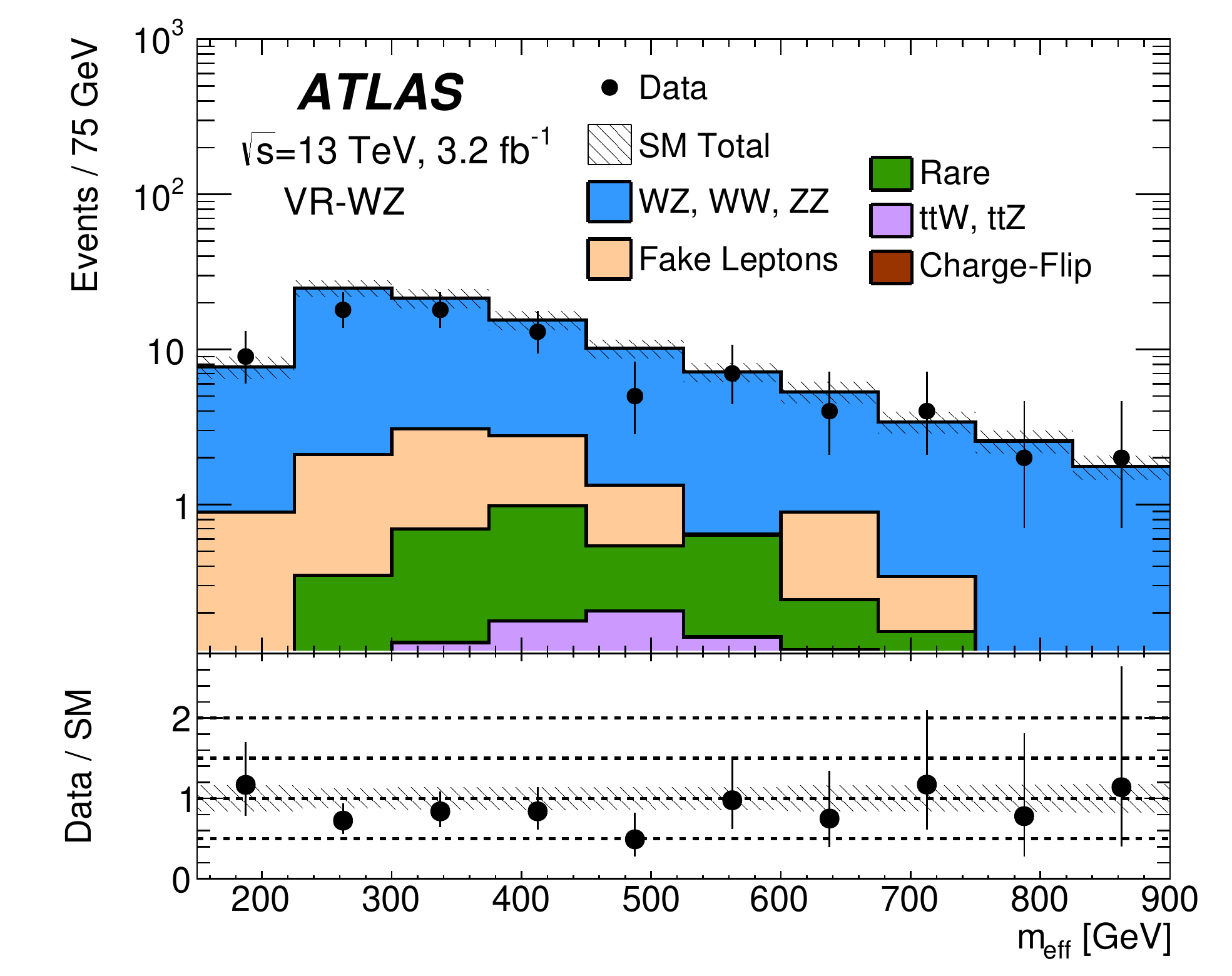}
\caption{}\label{fig:VRd}\end{subfigure}
\begin{subfigure}[t]{0.46\textwidth}\includegraphics[width=\textwidth]{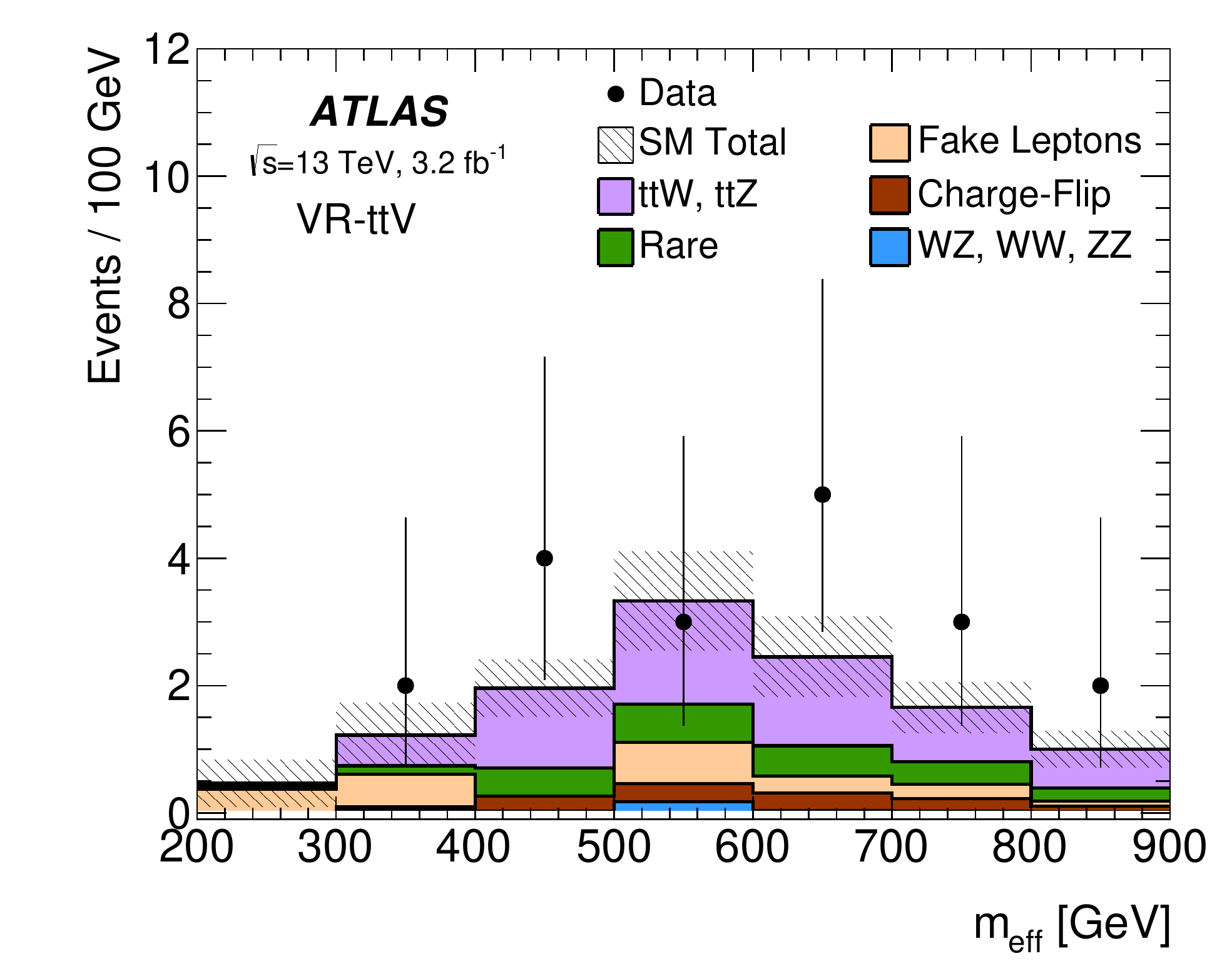}
\caption{}\label{fig:VRe}\end{subfigure}
\begin{subfigure}[t]{0.46\textwidth}\includegraphics[width=\textwidth]{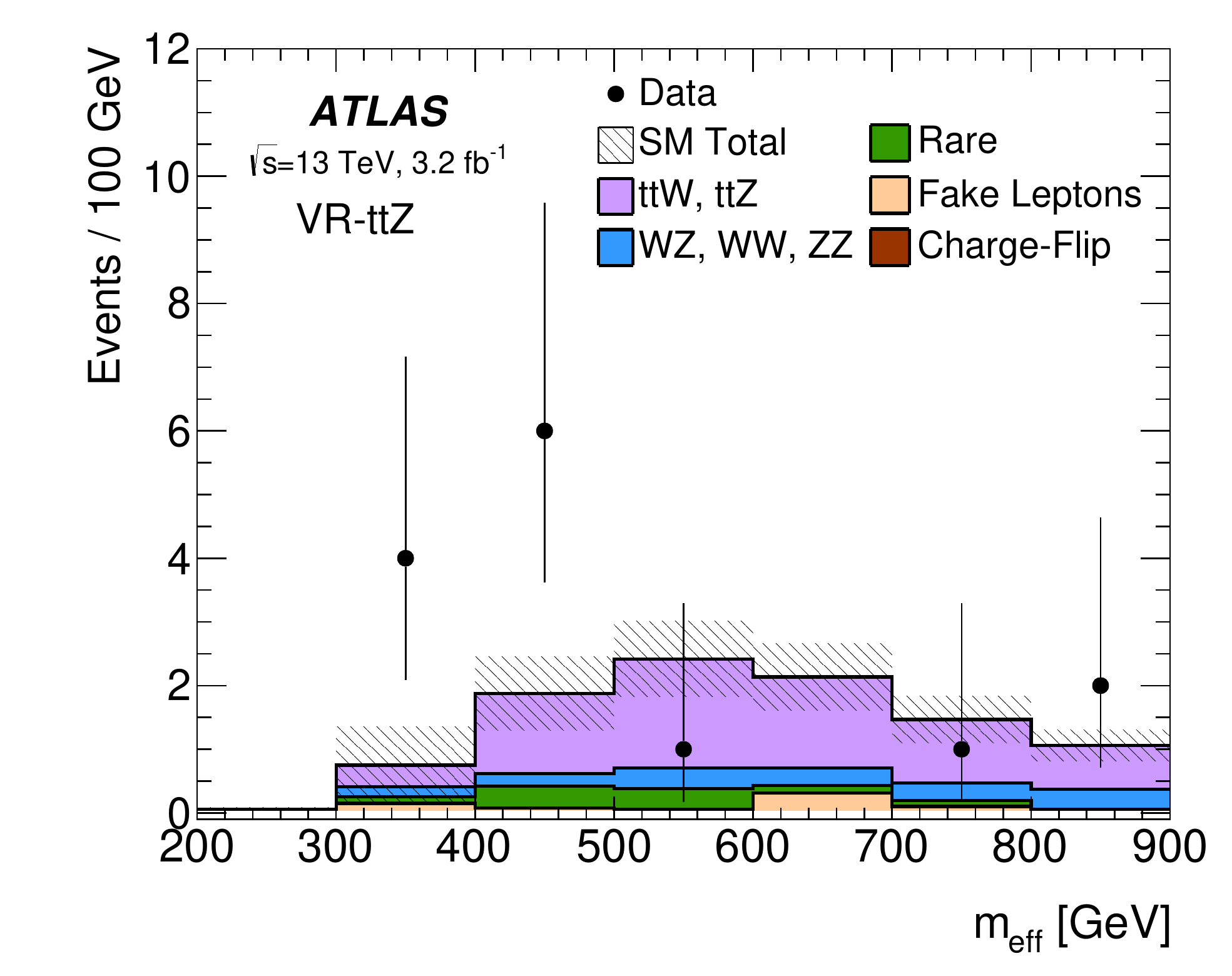}
\caption{}\label{fig:VRf}\end{subfigure}
\caption{Distributions of kinematic variables after a SS/3L selection 
including (a), (b) $\met>\SI{60}{GeV}$ and $N_\text{jet}^{25}\ge 2$, 
(c) a $b$-jet veto and $80<m_{\ell\ell}<\SI{100}{GeV}$, 
and (d), (e), (f) distributions in the validation regions. 
The statistical uncertainties in the background prediction are included in the uncertainty band, 
as well as the theory uncertainties for the backgrounds with prompt SS/3L, 
and the full systematic uncertainties for data-driven backgrounds. 
The last bin includes overflows. 
The ``Fake leptons'' category corresponds to FNP leptons (see text), 
and the ``Rare'' category contains the contributions from associated production of $\ttbar$ with $h/WW/t/\ttbar$, 
as well as $tZ$, $Wh$, $Zh$, and triboson production. 
The lower part of the figures (a)--(d) shows the ratio of data to the background prediction.}
\label{fig:Validation_plots} 
\end{figure}

To check the validity and robustness of the background estimates, 
the distributions of several discriminating variables in data are compared 
with the predicted background after various requirements on the number of jets and $b$-jets. 
Events are categorised based on the flavours of the selected leptons, and the different flavour channels are compared separately. 
Examples of such distributions are shown in Fig.~\ref{fig:VRa}--\ref{fig:VRc}
and illustrate that the predictions and data agree fairly well. 
The background estimates in a kinematic region close to the signal regions can also be observed in Fig.~\ref{fig:Results_SR_metD}, 
which shows the \met distributions in the signal regions before applying the \met requirements. 

\begin{table}[t!]
\caption{Summary of the event selection in the validation regions. 
Requirements are placed on the number of signal leptons ($N_{\rm{lept}}^{\rm{signal}}$) and candidate leptons ($N_{\rm{lept}}^{\rm{cand}}$), 
the number of jets with $\pt>\SI{25}{GeV}$ ($N_{\rm{jets}}^{25}$) or the number of $b$-jets with $\pt>\SI{20}{GeV}$ ($N_{b\rm{-jets}}^{20}$). 
The three leading-\pt leptons are referred to as $\ell_{1,2,3}$ with decreasing \pt and the two leading jets as $j_{1,2}$.
Additional requirements are set on the invariant mass of the two leading electrons $m_{ee}$, 
the presence of SS leptons or a pair of same-flavour opposite-sign leptons (SFOS) and its invariant mass $m_\text{SFOS}$. 
}
\hspace{0.5cm}
\def\arraystretch{1.1}
\label{tab:VRdef}
\centering
\resizebox{\textwidth}{!}
{\small
\begin{tabular}{c|c|c|c|c|c|l}
\hline 
\hline    
          &  $N_{\rm{lept}}^{\rm{signal}}$ ($N_{\rm{lept}}^{\rm{cand}}$)   & $N_{b\rm{-jets}}^{20}$  &  $N_{\rm{jets}}^{25}$  & \met\ [GeV] & \meff\ [GeV]  & Other \\
\hline\hline
VR-WW   & =2 (=2) &  =0  &  $\geq$2 & 35--200  & 300--900 & $m(j_1 j_2)>500$~GeV\\
          & =1 SS pair    &      &             &         &         & $\pt(j_2)>40$~GeV\\
          &      &      &             &         &         & $\pt(\ell_2)>30$~GeV\\
          &      &      &             &         &         & veto $80<m_{ee}<100$~GeV \\\hline
VR-WZ     & =3 (=3) &  =0  &  1--3     & 30--200  & $<$900 & $\pt(\ell_3)>30$~GeV \\ \hline
VR-ttV    &$\geq$2 (-) &$\geq$2&  $\geq$5 ($e^\pm e^\pm$,$e^\pm \mu^\pm$) & 20--200  & 200--900 & $\pt(\ell_2)>25$~GeV\\
          & $\geq$1 SS pair  &         &  $\geq$3 ($\mu^\pm \mu^\pm$)             &          &         & veto $\{\met>125$ and $\meff>\SI{650}{GeV}\}$\\ \hline
VR-ttZ    &$\geq$3 (-) & $\geq$1 & $\geq$4 ($=$1 $b$-jet)         & 20--150  & 100--900 & $\pt(\ell_2)>25$~GeV\\
          &$\geq$1 SFOS pair &         & $\geq$3 ($\geq$2 $b$-jets) &           &     & $\pt(\ell_3)>20$~GeV (if $e$)\\
          &        &         &                           &          &         & $80<m_\text{SFOS}<100$~GeV \\
\hline\hline
All VRs & \multicolumn{6}{c}{Veto events belonging to any SR, or if $\ell_1$ or $\ell_2$ is an electron with $|\eta|>1.37$ (except in VR-WZ)}\\
\hline\hline
\end{tabular}
}
\end{table}

Dedicated validation regions (VRs) are defined to test the estimate of the rare SM processes contributing to the signal regions, 
whose cross-sections have not yet been measured at $\sqrt s=13$ TeV. 
The corresponding selections are summarised in Table~\ref{tab:VRdef}. 
In these regions, upper bounds are placed on \met and \meff{} to reduce signal contamination, 
and the small residual overlap with the signal regions is resolved by vetoing events that contribute to the signal regions.  
To further reduce contributions from electron charge mis-identification, 
events are also vetoed if one of the two leading leptons is an electron with $|\eta|>1.37$, 
since contributions from charge-flip electrons are smaller in the central region due to the lower amount of crossed material. 
The purity of the targeted processes in these regions ranges from about 40\% to $80\%$. 
The VR-ttV and VR-ttZ regions overlap with each other, with 30\% of the $\ttbar V$ events in VR-ttV 
also present in VR-ttZ, 
and the contributions from the $\ttbar Z$ and $\ttbar W$ processes is similar in VR-ttV.

The observed yields in these validation regions, compared with the background predictions and uncertainties, 
can be seen in Table~\ref{tab:VR_yields}, and the effective mass distributions are shown in Fig.~\ref{fig:VRd}--\ref{fig:VRf}.
There is fair agreement between data and the estimated background for the validation regions, 
with the largest deviations being observed in VR-ttV with a 1.5$\sigma$ deviation.

\begin{table}
\caption{The numbers of observed data and expected background events for the validation regions. 
The ``Rare'' category contains the contributions from associated production of $\ttbar$ with $h/WW/t/\ttbar$, 
as well as $tZ$, $Wh$, $Zh$, and triboson production. 
Background categories shown as ``$-$'' denote that they cannot contribute to a given region (charge flips or $W^\pm W^\pm jj$ in 3-lepton regions). 
The individual uncertainties can be correlated and therefore do not necessarily add up in quadrature to the total systematic uncertainty. }
\label{tab:VR_yields}
\begin{center}
\setlength{\tabcolsep}{0.0pc}
{\small
\begin{tabular*}{\textwidth}{@{\extracolsep{\fill}}lcccc}
\noalign{\smallskip}\hline\hline\noalign{\smallskip}
          & VR-WW       & VR-WZ     & VR-ttV  & VR-ttZ     \\[-0.05cm]
\noalign{\smallskip}\hline\hline\noalign{\smallskip}
Observed events          & $4$       &  $82$   & $19$  & $14$ \\
\noalign{\smallskip}\hline\noalign{\smallskip}
Total background events & $3.4 \pm 0.8$ & $98 \pm 15$ & $12.1 \pm 2.7$\hspace*{1.75mm} & $9.7 \pm 2.5$\\
\noalign{\smallskip}\hline\noalign{\smallskip}
Fake/non-prompt leptons & $0.6 \pm 0.5$ & $8 \pm 6$ & $2.1 \pm 1.4$ & $0.6\pm 1.0$\\ 
Charge-flip & $0.26 \pm 0.05$ & $-$ & $1.14 \pm 0.15$ & $-$\\
$t\bar{t}W$ & $0.05 \pm 0.03$ & $0.25 \pm 0.09$ & $2.4 \pm 0.8$ & $0.10 \pm 0.03$\\
$t\bar{t}Z$ & $0.02 \pm 0.01$ & $0.72 \pm 0.26$ & $3.9 \pm 1.3$ & $6.3 \pm 2.1$\\
$WZ$ & $1.0 \pm 0.4$ & $78 \pm 13$ & $0.19 \pm 0.10$ & $1.2 \pm 0.4$\\
$W^\pm W^\pm jj$ & $1.3 \pm 0.5$ & $-$ & $0.02 \pm 0.03$ & $-$\\
$ZZ$ & $0.02 \pm 0.01$ & $8.2 \pm 2.8$ & $0.12 \pm 0.15$ & $0.30 \pm 0.19$\\
Rare & $0.10 \pm 0.05$ & $2.8 \pm 1.4$ & $2.3 \pm 1.2$ & $1.1 \pm 0.6$\\
\noalign{\smallskip}\hline\hline\noalign{\smallskip}
\end{tabular*}
}
\end{center}
\end{table}

%% file: results.tex
\begin{table}[htb!]
\begin{center}
\setlength{\tabcolsep}{0.0pc}
\caption{The number of observed data events and expected background contributions in the signal regions. 
The $p$-value of the observed events for the background-only hypothesis is denoted by $p(s = 0)$. 
The ``Rare'' category contains the contributions from associated production of $\ttbar$ with $h/WW/t/\ttbar$, 
as well as $tZ$, $Wh$, $Zh$, and triboson production. 
Background categories shown as ``$-$'' denote that they cannot contribute to a given region (charge flips or $W^\pm W^\pm jj$ in 3-lepton regions). 
The individual uncertainties can be correlated and therefore do not necessarily add up in quadrature to the total systematic uncertainty. 
}
\label{tab:SR_yields}
{\small
\begin{tabular*}{\textwidth}{@{\extracolsep{\fill}}lcccc}
\noalign{\smallskip}\hline\hline\noalign{\smallskip}
         & SR0b3j         & SR0b5j     & SR1b & SR3b     \\[-0.05cm]
\noalign{\smallskip}\hline\hline\noalign{\smallskip}
Observed events         & $3$     &  $3$  & $7$  & $1$            \\
\noalign{\smallskip}\hline\noalign{\smallskip}
Total background events & $1.5 \pm 0.4$ & $0.88 \pm 0.29$ & $4.5 \pm 1.0$ & $0.80 \pm 0.25$\\
$p(s = 0)$                &  0.13  &  0.04  &  0.15  &   0.36   \\
\noalign{\smallskip}\hline\noalign{\smallskip}
Fake/non-prompt leptons & $<0.2$ & $0.05\pm 0.18$ & $0.8 \pm 0.8$ & $0.13 \pm 0.17$\\
Charge-flip & $-$ & $0.02 \pm 0.01$ & $0.60 \pm 0.12$ & $0.19 \pm 0.06$\\
$t\bar{t}W$ & $0.02 \pm 0.01$ & $0.08 \pm 0.04$ & $1.1 \pm 0.4$ & $0.10 \pm 0.05$\\
$t\bar{t}Z$ & $0.10 \pm 0.04$ & $0.05 \pm 0.03$ & $0.92 \pm 0.31$ & $0.14 \pm 0.06$\\
$WZ$ & $1.2 \pm 0.4$ & $0.48 \pm 0.20$ & $0.18 \pm 0.11$ & $<0.02$\\
$W^\pm W^\pm jj$ & $-$ & $0.12 \pm 0.07$ & $0.03 \pm 0.02$ & $<0.01$\\
$ZZ$ & $<0.03$ & $<0.04$ & $<0.03$ & $<0.03$\\
Rare & $0.14 \pm 0.08$ & $0.07 \pm 0.05$ & $0.8 \pm 0.4$ & $0.24 \pm 0.14$\\  
\noalign{\smallskip}\hline\hline\noalign{\smallskip}
\end{tabular*}
}
\end{center}
\end{table}

\begin{figure}[t!]
\centering
\begin{subfigure}[t]{0.49\textwidth}\includegraphics[width=\textwidth]{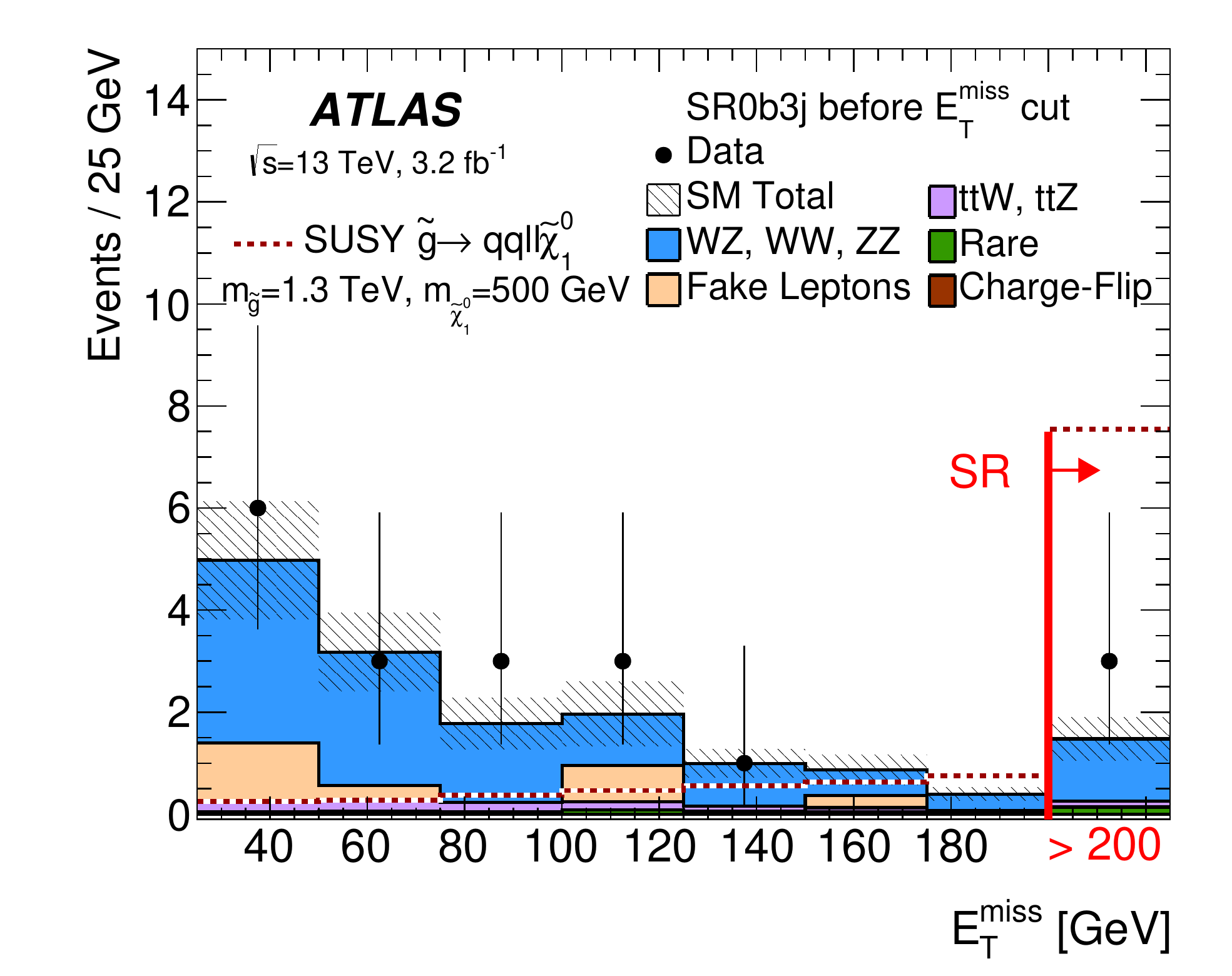}
\caption{}\label{fig:Results_SR0b3j}\end{subfigure}
\begin{subfigure}[t]{0.49\textwidth}\includegraphics[width=\textwidth]{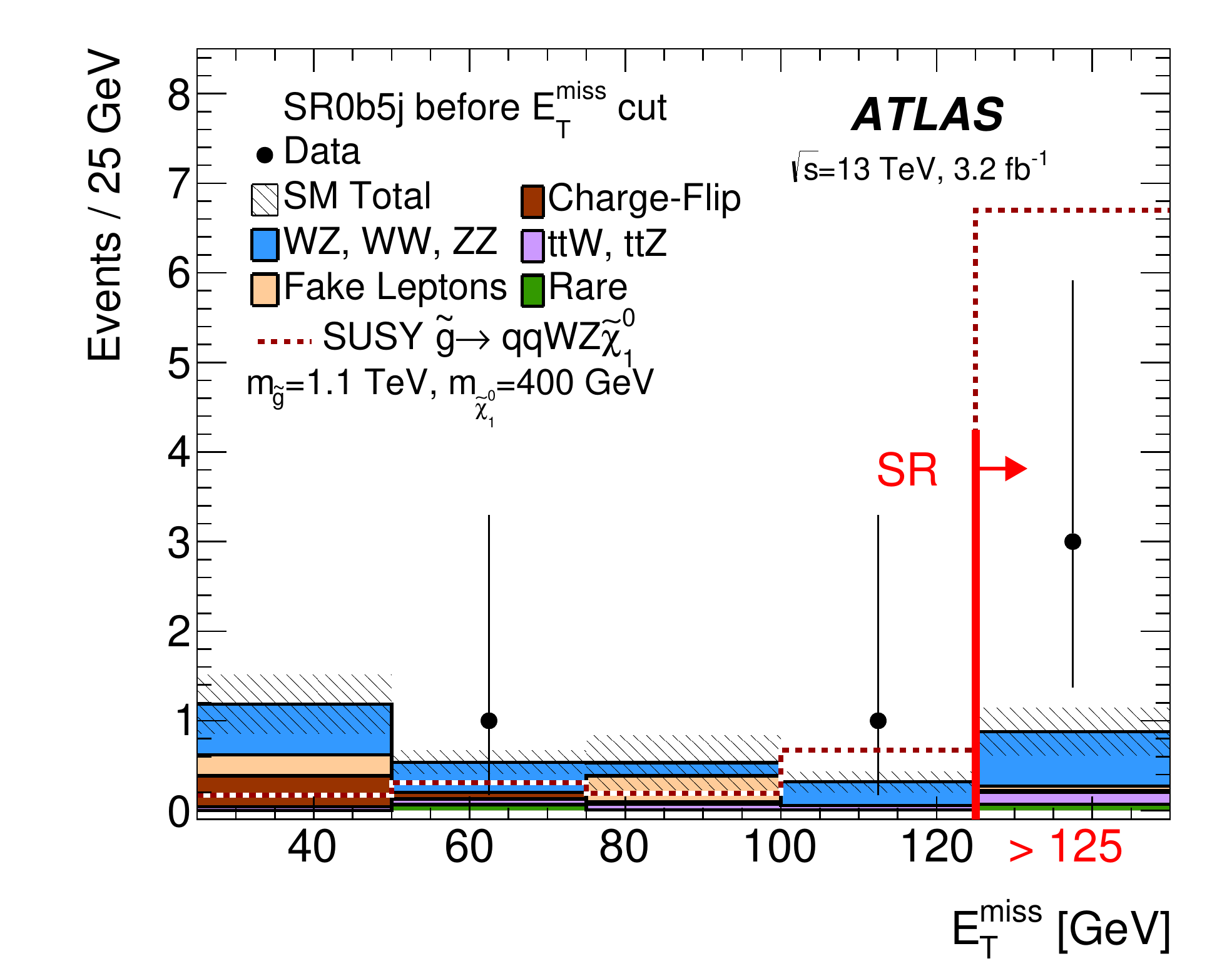}
\caption{}\label{fig:Results_SR0b5j}\end{subfigure}
\begin{subfigure}[t]{0.49\textwidth}\includegraphics[width=\textwidth]{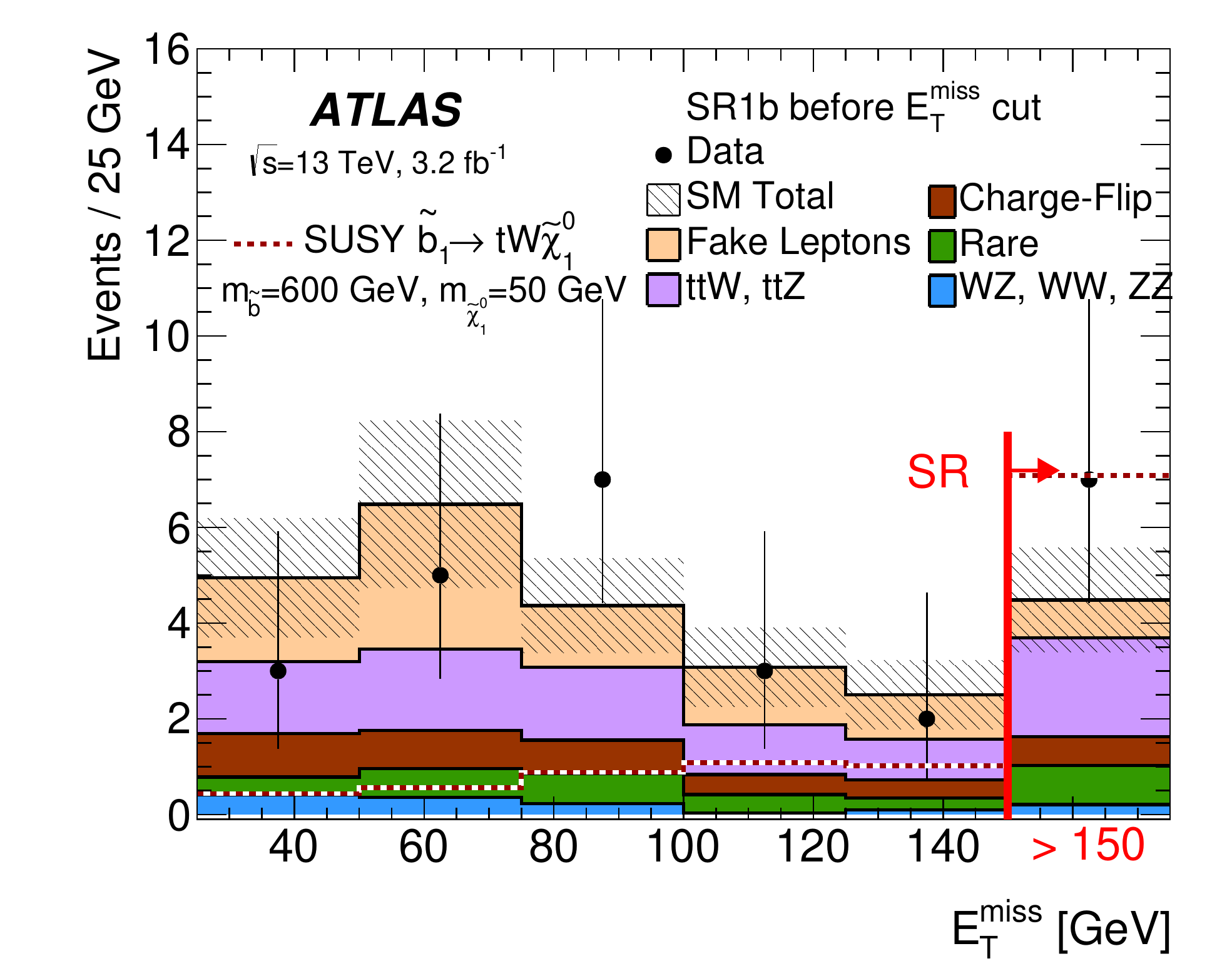}
\caption{}\label{fig:Results_SR1b}\end{subfigure}
\begin{subfigure}[t]{0.49\textwidth}\includegraphics[width=\textwidth]{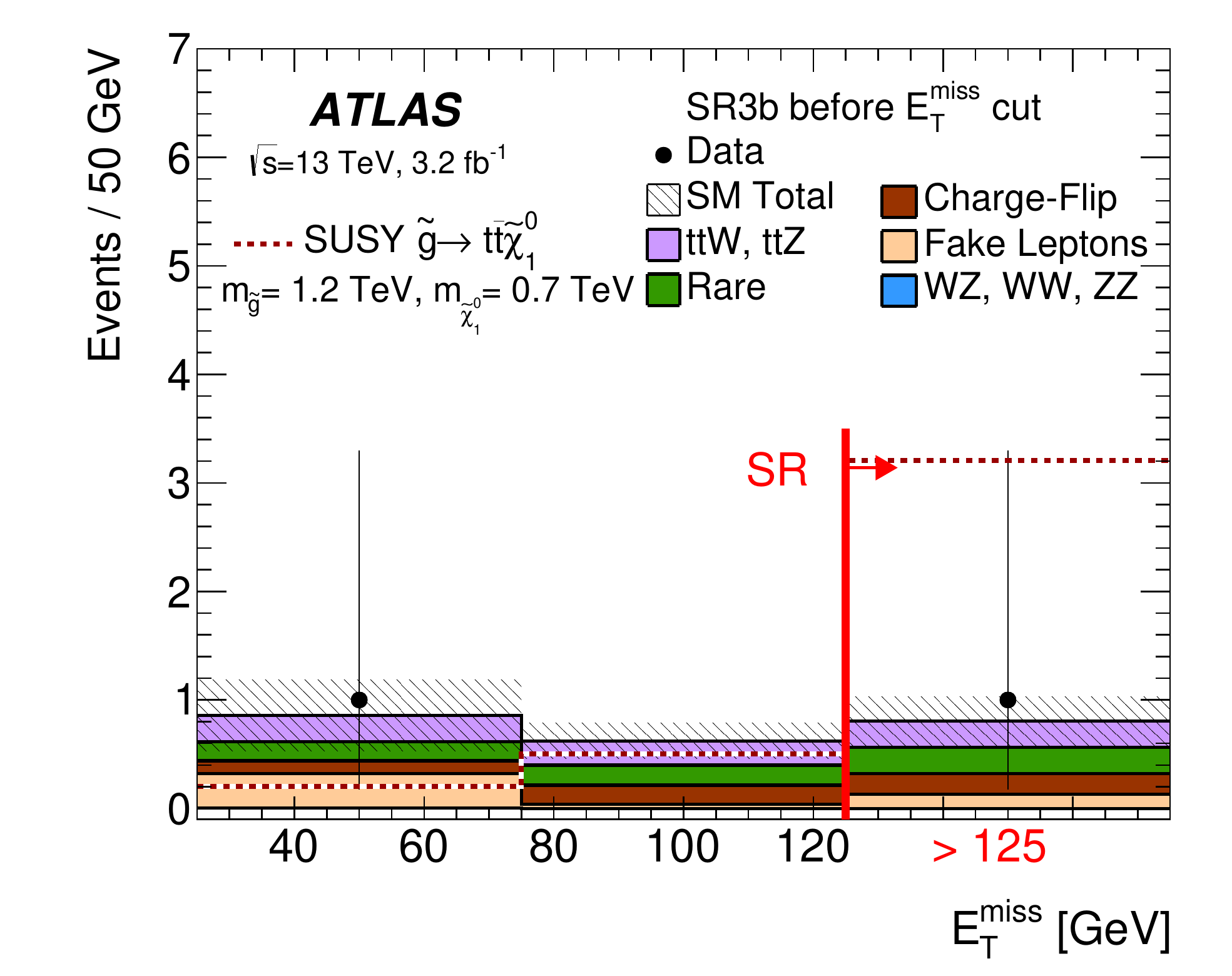}
\caption{}\label{fig:Results_SR3b}\end{subfigure}
\caption{Missing transverse momentum distributions after (a) SR0b3j, (b) SR0b5j, (c) SR1b and (d) SR3b selection, beside the \met requirement. 
The results in the signal regions are shown in the last (inclusive) bin of each plot. 
The statistical uncertainties in the background prediction are included in the uncertainty band, 
as well as the theory uncertainties for the backgrounds with prompt SS/3L, 
and the full systematic uncertainties for data-driven backgrounds. 
The ``Fake leptons'' category corresponds to FNP leptons (see text), 
and the ``Rare'' category contains the contributions from associated production of $\ttbar$ with $h/WW/t/\ttbar$, 
as well as $tZ$, $Wh$, $Zh$, and triboson production. 
}
\label{fig:Results_SR_metD} 
\end{figure}

Figure~\ref{fig:Results_SR_metD} shows the data \met distributions after the signal region selections (beside that on \met) in data 
together with the expected contributions from all the SM backgrounds with their total statistical and systematic uncertainties. 
For illustration, a typical SUSY signal distribution corresponding to the most relevant benchmark scenario 
in each SR is displayed.
The detailed yields for data and the different sources of SM background in the signal regions 
are presented in Table~\ref{tab:SR_yields}. 
The uncertainties amount to 22--34\% of the total background depending on the signal region. 
In all four SRs the number of data events exceeds the expectation but is consistent within the uncertainties, 
the smallest $p$-value for the SM-only hypothesis being 0.04 for SR0b5j. 
Out of the 14 events in the SRs, 2 of the events in SR1b and the 3 events in SR0b3j contain three leptons. 
None of those events contain three leptons of equal charge, or are present in more than one SR.

In the absence of any significant deviations from the SM predictions, upper limits on possible BSM contributions to the signal regions are computed, 
in particular in the context of the SUSY benchmark scenarios described in Section~\ref{sec:intro}. 
The HistFitter framework~\cite{Baak:2014wma}, which utilises a profile-likelihood-ratio test~\cite{Cowan:2010js}, 
is used to establish 95\% confidence intervals using the CL$_\mathrm{s}$ prescription~\cite{Read_CLs}. 
The likelihood is built as the product of a Poisson probability density function describing the observed number of events in the signal region 
and Gaussian distributions constraining the nuisance parameters 
associated with the systematic uncertainties whose widths correspond to the sizes of these uncertainties; 
Poisson distributions are used instead for MC statistical uncertainties. 
Correlations of a given nuisance parameter across the different sources of backgrounds and the signal are taken into account when relevant. 
The statistical tests are performed independently for each of the signal regions. 

Table~\ref{tab:upperlimits} presents 95\% confidence level (CL) model-independent upper limits 
on the number of BSM events, $N_\mathrm{BSM}$, that may contribute to the signal regions. 
Normalising these by the integrated luminosity $L$ of the data sample, they can be interpreted as upper limits on the visible BSM cross-section $\sigma_{\rm{vis}}$, 
defined as the product $\sigma_{\rm{prod}}\times A \times\epsilon=N_\mathrm{BSM}/L$ of production cross-section, acceptance and reconstruction efficiency. 

\begin{table}[htb!]
\centering
\caption{Signal model-independent upper limits on the number of BSM events ($N_{\rm{BSM}}$) 
  and the visible signal cross-section ($\sigma_{\rm{vis}}$) in the four SRs. 
  The numbers (in parentheses) give the observed (expected under the SM hypothesis) 95\% CL upper
  limits. Calculations are performed with pseudo-experiments.
  The $\pm$1$\sigma$ variations on the expected limit due to the statistical and systematic uncertainties in the background prediction are also shown. 
}
\label{tab:upperlimits}
{\small
\renewcommand{\arraystretch}{1.4}
\begin{tabular*}{\textwidth}{@{\extracolsep{\fill}}lrrrr}
\noalign{\smallskip}\hline\hline\noalign{\smallskip}
         & SR0b3j         & SR0b5j     & SR1b & SR3b     \\[-0.05cm]
\noalign{\smallskip}\hline\hline\noalign{\smallskip}
$N_{\rm{BSM}}^{\rm{obs}}$ ($N_{\rm{BSM}}^{\rm{exp}}$) 
 & $5.9$  $({4.1}^{+1.6}_{-0.8})$
 & $6.4$ $({3.6}^{+1.2}_{-1.1})$
 & $8.8$ $({6.0}^{+2.6}_{-1.6})$ 
 & $3.8$ $({3.7}^{+1.1}_{-0.5})$ \\
$\sigma_{\rm{vis}}^{\rm{obs}}$ [fb] & 1.8 & 2.0 & 2.8 & 1.2\\
\noalign{\smallskip}\hline\hline\noalign{\smallskip}
  \end{tabular*}
}
\end{table} 

\begin{figure}[htb!]
\centering
\begin{subfigure}[t]{0.49\textwidth}\includegraphics[width=\textwidth]{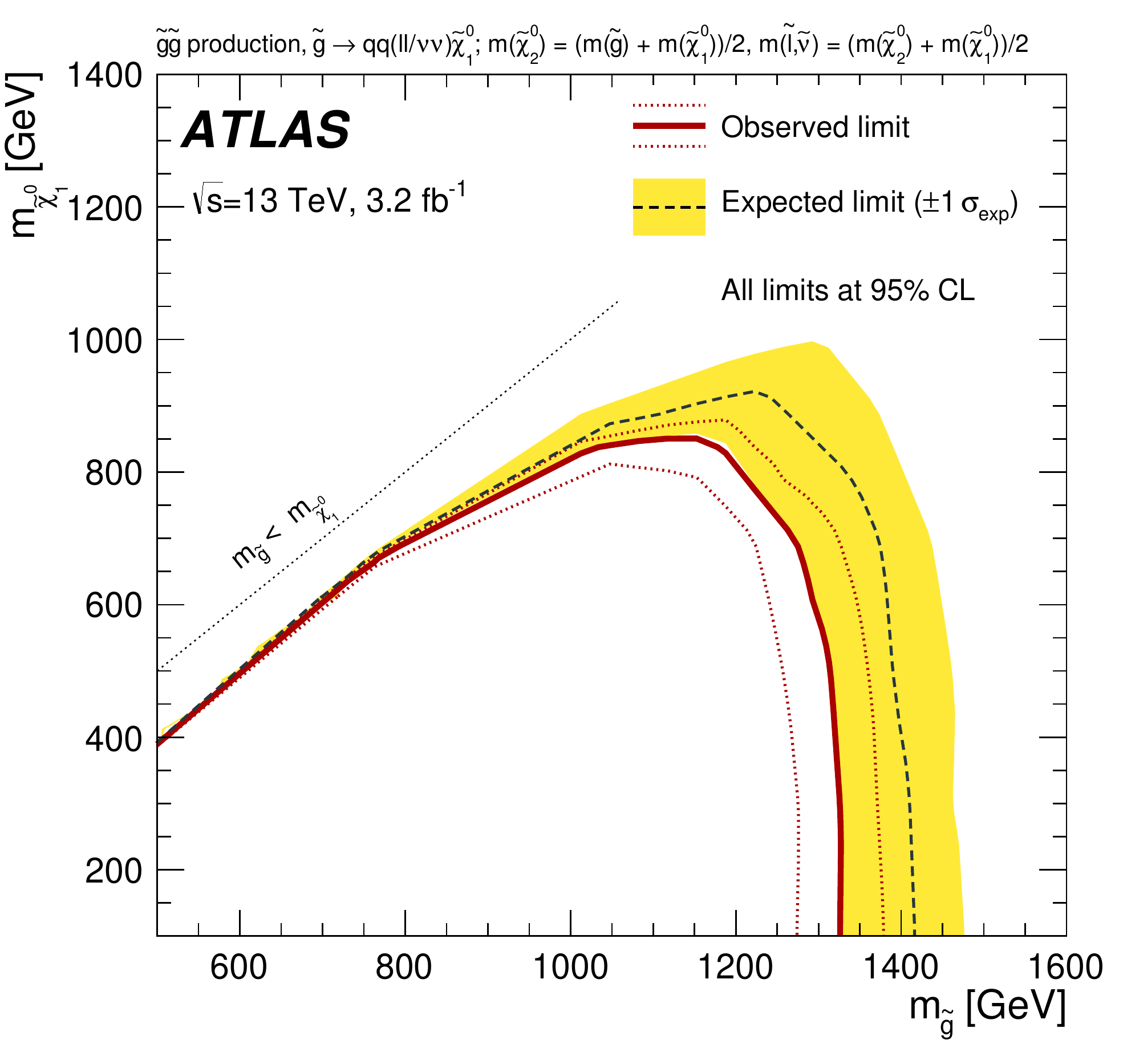}
\caption{$\gluino\to q\bar q \ell\ell\ninoone$ scenario, SR0b3j}\label{fig:limits_SR0b3j}\end{subfigure}
\begin{subfigure}[t]{0.49\textwidth}\includegraphics[width=\textwidth]{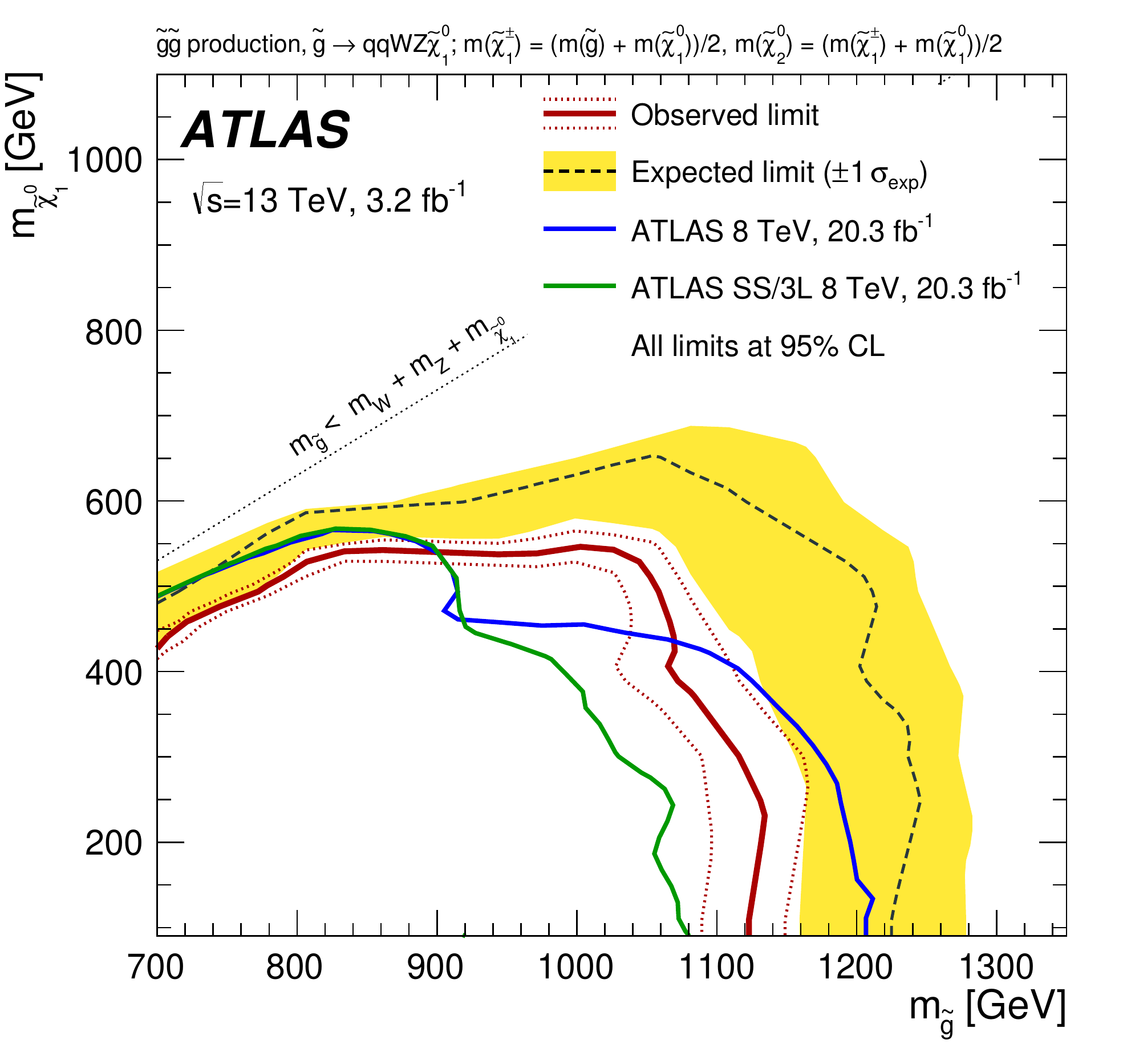}
\caption{$\gluino\to q\bar q' WZ\ninoone$ scenario, SR0b5j}\label{fig:limits_SR0b5j}\end{subfigure}
\par\bigskip
\begin{subfigure}[t]{0.49\textwidth}\includegraphics[width=\textwidth]{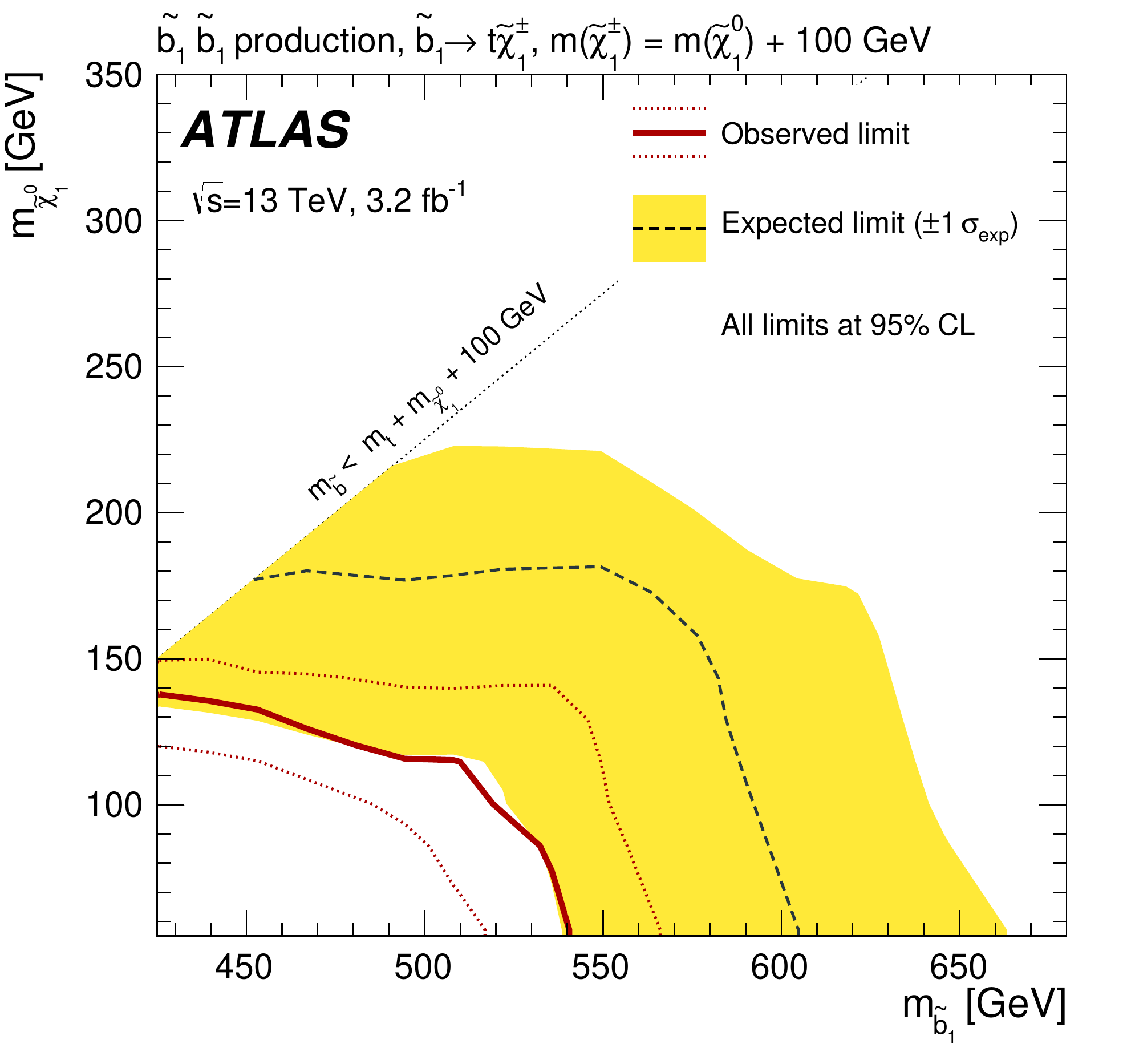}
\caption{$\sbottomone\to tW^-\ninoone$ scenario, SR1b}\label{fig:limits_SR1b}\end{subfigure}
\begin{subfigure}[t]{0.49\textwidth}\includegraphics[width=\textwidth]{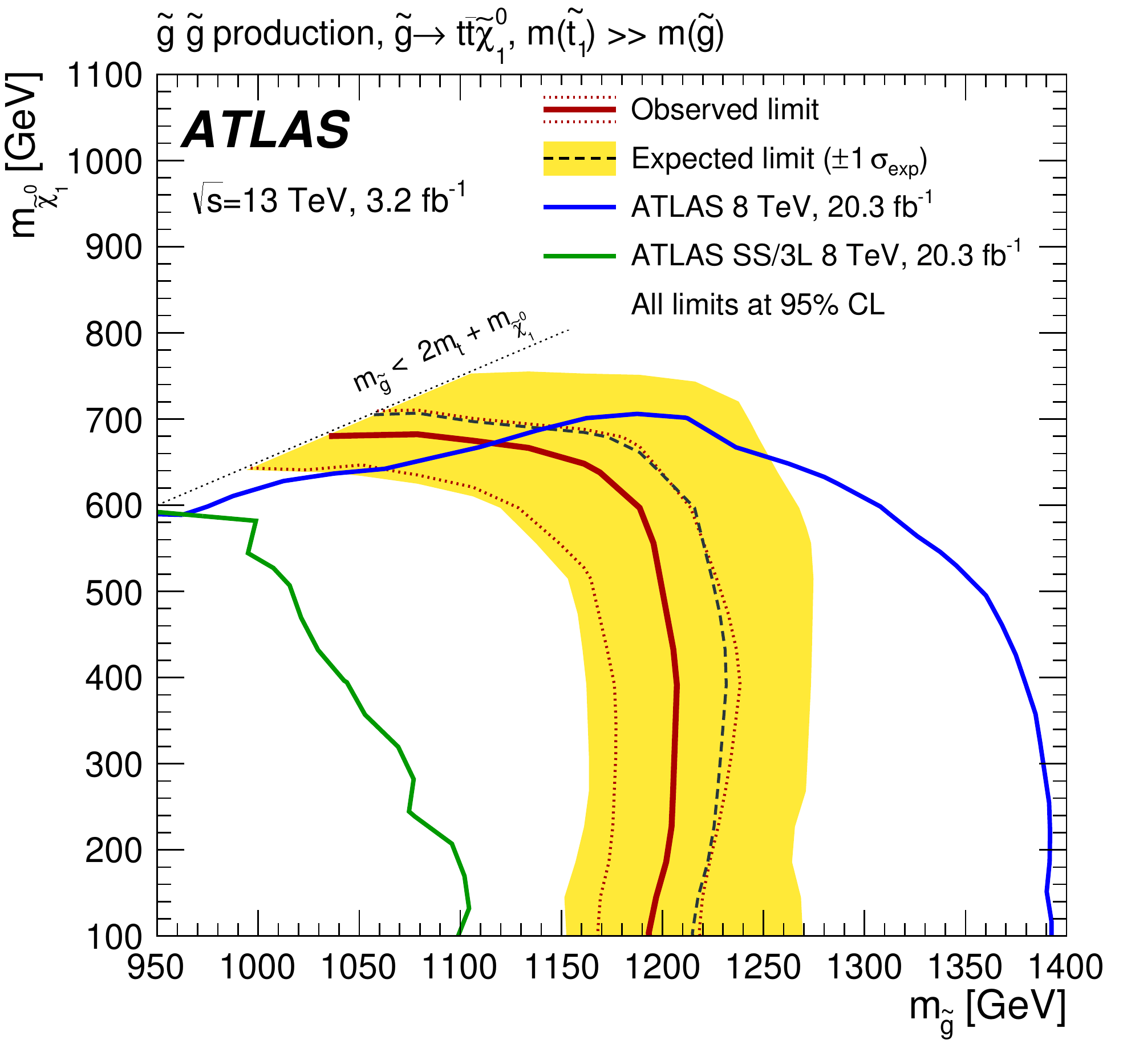}
\caption{$\gluino\to t\bar t\ninoone$ scenario, SR3b}\label{fig:limits_SR3b}\end{subfigure}
\caption{
Observed and expected exclusion limits on the \gluino, \sbottomone and \ninoone masses 
in the context of SUSY scenarios with simplified mass spectra 
featuring $\gluino\gluino$ or $\sbottomone\sbottomonebar$ pair production with exclusive decay modes. 
The signal region used to obtain the limits is specified for each scenario. 
The contours of the band around the expected limit are the $\pm$1$\sigma$ results, 
  including all uncertainties except theoretical uncertainties on the signal cross-section. The dotted lines around the observed
    limit illustrate the change in the observed limit as the nominal signal cross-section is scaled up and down
    by the theoretical uncertainty. All limits are computed at 95\% CL. 
    The diagonal lines indicate the kinematic limit for the decays in each specified scenario.  
For figures (b) and (d), results are compared with the observed limits obtained by previous ATLAS searches~\cite{paperSS3L,Aad:2015iea,Aad:2015pfx}. 
For figures (a) and (c), a direct comparison with earlier searches is not possible, due to differing model assumptions. 
}
\label{fig:Results_Limits} 
\end{figure} 

Exclusion limits are also set on the masses of the superpartners involved in the four SUSY benchmark scenarios considered in this analysis. 
Simplified models corresponding to a single production mode and with 100\% branching ratio to a specific decay chain are used, 
with the masses of the SUSY particles not involved in the process set to very high values. 
Figure~\ref{fig:Results_Limits} shows the limits 
on the mass of the $\ninoone$ as a function of the $\gluino$ or $\sbottomone$ mass. 
In some cases, the new limits set by this analysis can be compared 
with the existing limits set by the combination of ATLAS SUSY searches with 8 TeV data~\cite{Aad:2015iea,Aad:2015pfx}. 
For parts of the parameter space, the sensitivity reached with the 13 TeV dataset exceeds that of the 8 TeV dataset,
and additional parameter space regions can be excluded, especially for large neutralino masses. 

Signal models featuring gluino pair production with a subsequent gluino decay via $\ninotwo$ and light sleptons\\ 
($\gluino\to q\bar q\ninotwo\to q\bar q (\ell\slepton^*/\nu\tilde{\nu}^*)\to q\bar q(\ell^+\ell^-/\nu\nu)\ninoone$) 
are probed using SR0b3j (Fig.~\ref{fig:limits_SR0b3j}).
In this simplified model, the gluinos decay into $u\bar u$, $d\bar d$, $s\bar s$ or $c\bar c$ with equal probabilities, 
and the six types of leptons are also produced in the $\tilde\chi_2^0$ decays with equal probabilities. 
The $\ninotwo$ mass is set to $m_{\ninotwo}=(m_{\gluino} + m_{\ninoone})/2$, 
with the $\slepton$ and $\tilde{\nu}$ masses set to $m_{\slepton,\tilde{\nu}}=(m_{\ninotwo} + m_{\ninoone})/2$.
Gluino masses up to $m_{\gluino}\approx\SI{1.3}{TeV}$ for a light \ninoone and \ninoone masses up to $m_{\ninoone}\approx\SI{850}{GeV}$ for gluinos with $m_{\gluino}\approx\SI{1.1}{TeV}$ are excluded in this scenario. 

Similarly, models with gluino production  with a subsequent two-step gluino decay via $\chinoonepm$ and $\ninotwo$\\ 
($\gluino\to q\bar q \chinoonepm \to q\bar q W\ninotwo \to  q\bar q W Z \ninoone$) 
are probed with SR0b5j (Fig.~\ref{fig:limits_SR0b5j}).
In this simplified model, the gluinos decay into $u\bar u$, $d\bar d$, $s\bar s$ or $c\bar c$ with equal probabilities. 
The $\chinoonepm$ mass is set to $m_{\chinoonepm}=(m_{\gluino} + m_{\ninoone})/2$ and
the $\ninotwo$ mass is set to $m_{\ninotwo}=(m_{\chinoonepm} + m_{\ninoone})/2$; 
$W$ and $Z$ bosons produced in the decay chain are not necessarily on-shell. 
The exclusion limits in this scenario reach $m_{\gluino}\approx\SI{1.1}{TeV}$ (for light $\ninoone$) and $m_{\ninoone}\approx\SI{550}{GeV}$ (for $m_{\gluino}\approx\SI{1.0}{TeV}$).

Exclusion limits in a simplified model of bottom squark production with chargino-mediated $\sbottomone\to tW^-\ninoone$ decays are 
obtained with SR1b (Fig.~\ref{fig:limits_SR1b}).
In this model the $\chinoonepm$ mass is set to $m_{\chinoonepm}=m_{\ninoone} + \SI{100}{GeV}$.
The limits can reach mass values of $m_{\sbottomone}\approx\SI{540}{GeV}$ for a light $\ninoone$, 
while $m_{\ninoone}\lesssim\SI{140}{GeV}$ are also excluded for $m_{\sbottomone}\approx\SI{425}{GeV}$, 
significantly extending the previous limits obtained at $\sqrt{s}=8$~TeV~\cite{Aad:2015pfx} 
which excluded $m_{\sbottomone}\lesssim\SI{470}{GeV}$ for $m_{\ninoone}\approx\SI{60}{GeV}$ for a similar model.

Finally, SR3b is used to set limits on masses in a simplified model with 
gluino pair production and $\gluino\to t\bar t\ninoone$ decays 
via an off-shell top squark (Fig.~\ref{fig:limits_SR3b}). 
In that case, gluino masses of $m_{\gluino}\lesssim\SI{1.2}{TeV}$ are excluded for $m_{\ninoone}\lesssim\SI{600}{GeV}$, 
and $\ninoone$ masses up to $m_{\ninoone}\approx\SI{680}{GeV}$ are also excluded for $m_{\gluino}\approx\SI{1.05}{TeV}$.

%% file: Acknowledgements.tex

We thank CERN for the very successful operation of the LHC, as well as the
support staff from our institutions without whom ATLAS could not be
operated efficiently.

We acknowledge the support of ANPCyT, Argentina; YerPhI, Armenia; ARC, Australia; BMWFW and FWF, Austria; ANAS, Azerbaijan; SSTC, Belarus; CNPq and FAPESP, Brazil; NSERC, NRC and CFI, Canada; CERN; CONICYT, Chile; CAS, MOST and NSFC, China; COLCIENCIAS, Colombia; MSMT CR, MPO CR and VSC CR, Czech Republic; DNRF and DNSRC, Denmark; IN2P3-CNRS, CEA-DSM/IRFU, France; GNSF, Georgia; BMBF, HGF, and MPG, Germany; GSRT, Greece; RGC, Hong Kong SAR, China; ISF, I-CORE and Benoziyo Center, Israel; INFN, Italy; MEXT and JSPS, Japan; CNRST, Morocco; FOM and NWO, Netherlands; RCN, Norway; MNiSW and NCN, Poland; FCT, Portugal; MNE/IFA, Romania; MES of Russia and NRC KI, Russian Federation; JINR; MESTD, Serbia; MSSR, Slovakia; ARRS and MIZ\v{S}, Slovenia; DST/NRF, South Africa; MINECO, Spain; SRC and Wallenberg Foundation, Sweden; SERI, SNSF and Cantons of Bern and Geneva, Switzerland; MOST, Taiwan; TAEK, Turkey; STFC, United Kingdom; DOE and NSF, United States of America. In addition, individual groups and members have received support from BCKDF, the Canada Council, CANARIE, CRC, Compute Canada, FQRNT, and the Ontario Innovation Trust, Canada; EPLANET, ERC, FP7, Horizon 2020 and Marie Sk{\l}odowska-Curie Actions, European Union; Investissements d'Avenir Labex and Idex, ANR, R{\'e}gion Auvergne and Fondation Partager le Savoir, France; DFG and AvH Foundation, Germany; Herakleitos, Thales and Aristeia programmes co-financed by EU-ESF and the Greek NSRF; BSF, GIF and Minerva, Israel; BRF, Norway; the Royal Society and Leverhulme Trust, United Kingdom.

The crucial computing support from all WLCG partners is acknowledged
gratefully, in particular from CERN and the ATLAS Tier-1 facilities at
TRIUMF (Canada), NDGF (Denmark, Norway, Sweden), CC-IN2P3 (France),
KIT/GridKA (Germany), INFN-CNAF (Italy), NL-T1 (Netherlands), PIC (Spain),
ASGC (Taiwan), RAL (UK) and BNL (USA) and in the Tier-2 facilities
worldwide.

%% file: atlas_authlist.tex
\begin{flushleft}
{\Large The ATLAS Collaboration}

\bigskip

G.~Aad$^\textrm{\scriptsize 86}$,
B.~Abbott$^\textrm{\scriptsize 113}$,
J.~Abdallah$^\textrm{\scriptsize 151}$,
O.~Abdinov$^\textrm{\scriptsize 11}$,
B.~Abeloos$^\textrm{\scriptsize 117}$,
R.~Aben$^\textrm{\scriptsize 107}$,
M.~Abolins$^\textrm{\scriptsize 91}$,
O.S.~AbouZeid$^\textrm{\scriptsize 137}$,
H.~Abramowicz$^\textrm{\scriptsize 153}$,
H.~Abreu$^\textrm{\scriptsize 152}$
% ...
\bigskip
\\
% ...
$^{11}$ Institute of Physics, Azerbaijan Academy of Sciences, Baku, Azerbaijan\\
% ...
$^{86}$ CPPM, Aix-Marseille Universit{\'e} and CNRS/IN2P3, Marseille, France\\
% ...
$^{91}$ Department of Physics and Astronomy, Michigan State University, East Lansing MI, United States of America\\
% ...
$^{107}$ Nikhef National Institute for Subatomic Physics and University of Amsterdam, Amsterdam, Netherlands\\
% ...
$^{113}$ Homer L. Dodge Department of Physics and Astronomy, University of Oklahoma, Norman OK, United States of America\\
% ...
$^{117}$ LAL, Univ. Paris-Sud, CNRS/IN2P3, Universit{\'e} Paris-Saclay, Orsay, France\\
% ...
$^{137}$ Santa Cruz Institute for Particle Physics, University of California Santa Cruz, Santa Cruz CA, United States of America\\
% ...
$^{151}$ Institute of Physics, Academia Sinica, Taipei, Taiwan\\
$^{152}$ Department of Physics, Technion: Israel Institute of Technology, Haifa, Israel\\
$^{153}$ Raymond and Beverly Sackler School of Physics and Astronomy, Tel Aviv University, Tel Aviv, Israel\\
% ...
\end{flushleft}